\documentclass{aa} 
\usepackage{graphicx} 
\usepackage{txfonts}
\usepackage{subcaption}
\usepackage{placeins}
\usepackage{hyperref}
\usepackage{color}
\usepackage{booktabs}
\usepackage{xcolor}
\usepackage{enumitem}
\usepackage{array}
\usepackage{tabularx}
\usepackage{geometry}
\usepackage{titlesec}
\usepackage{multicol}
\usepackage{float}

\newcommand{\blob}{G319.8-2.0}
\newcommand{\GG}{G320.2-3.6}

\begin{document}

\title{Multiwavelength study of supernova remnants in the Circinus constellation with SRG/eROSITA}

\titlerunning{SNRs in Circinus}

  \author{A. Khokhriakova
          \inst{1}
          \and
          W. Becker\inst{1,2}
          }

  \institute{
              Max-Planck Institut für extraterrestrische Physik, Giessenbachstraße, 85748 Garching, Germany
        \and
              Max-Planck Institut für Radioastronomie, Auf dem Hügel 69, 53121 Bonn, Germany
        \email{alena@mpe.mpg.de}
             }
             
    \date{Received XXX; accepted YYY}

\abstract{
{Supernova remnants (SNRs) are key tracers of stellar feedback and the chemical and dynamical evolution of the interstellar medium. However, identifying SNRs in complex regions remains challenging, particularly when only archival multi-wavelength data are available.}
%
{We aim to search for previously unrecognized SNR candidates in the Circinus constellation by combining optical imaging, radio continuum data, and available SRG/eROSITA all-sky survey X-ray observations, to provide a prioritized list of targets for future confirmation. Candidate remnants were classified according to the strength of the combined diagnostics.}
%
{This joint analysis yields improved constraints on the morphology and
spectral properties of known remnants in the region, including MSH 15-52 and G320.6-1.6. In addition, we identify a new promising SNR candidate, G320.2-3.6.}
%
{Our study provides a first multi-wavelength search of SNR candidates in the Circinus region. In addition to G320.2-3.6, three objects (G319.8-2.0, G321.8-1.1, and G320.5-1.8) emerge as promising targets for follow-up radio,  spectroscopic optical, and deeper, spatially resolved X-ray observations to  confirm their nature.}

 }

   \keywords{
              ISM: individual objects: MSH 15-52 -
              ISM: individual objects: G320.6-1.6 - 
              ISM: individual objects: G320.2-3.6 -
              ISM: individual objects: G319.8-2.0 -
              ISM: individual objects: G321.9-1.1 -
              ISM: individual objects: G320.5-1.8
               }

   \maketitle

\section{Introduction}

Supernova remnants (SNRs) are the expanding structures left behind after a supernova explosion, marking the final stages of stellar evolution of massive stars or white dwarfs in binary systems. These remnants are crucial for studying high-energy processes, stellar feedback, and the interstellar medium (ISM) as they disperse heavy elements synthesized during stellar lifetimes and the explosion itself, enriching the surrounding medium for the next generation of star formation. Additionally, SNRs are considered major accelerators of cosmic rays \citep{2013Sci...339..807A, 2010ApJ...718...31P}, as their expanding shock fronts can energize charged particles to relativistic speeds through mechanisms like diffusive shock acceleration \citep{1978MNRAS.182..147B}. 

Despite their significance, the Galactic population of SNRs remains not completely understood. A long-standing issue is the discrepancy between the number of observed SNRs and theoretical expectations (the problem of missing SNRs). 
As of October 2025, the Green catalogue\footnote{Available at \url{https://www.mrao.cam.ac.uk/surveys/snrs/}} \citep{2025JApA...46...14G} lists only 310 confirmed SNRs, a number significantly lower than the $\gtrsim 1000$ predicted from simple considerations of the Galactic supernova rate and typical SNR lifetimes \citep{1991ApJ...378...93L, 1994ApJS...92..487T}.
A recent work \cite{2022ApJ...940...63R} estimates the number of SNRs in our Galaxy to be in the range $\sim 2400$ to $5600$. 
This discrepancy is generally attributed to the observational selection effects, in particular the bias against detecting evolved, low surface-brightness remnants \citep{2021MNRAS.504.1536V}, leading to an incomplete and biased census of the Galactic SNR population.

SNRs emit across the entire electromagnetic spectrum, from radio to gamma-rays. Each band probes different physical components of the remnant. The detectability of a remnant in any particular band depends on its evolutionary state, environment, and the amount of foreground absorption.

In the Milky Way, the majority of known remnants have been identified through their radio emission (see e.g. \citealt{2025JApA...46...14G} and references therein). 
This is because SNRs remain radio sources during their whole lifetime, and radio wavelengths are largely unaffected by dust extinction, which is severe in the Galactic plane where most SNRs reside.
Combined with the availability of wide-area radio surveys, this makes the radio channel an efficient waveband for the discovery of SNRs in the Milky Way and other neighbouring Galaxies \citep{2010MNRAS.407.1301B,2005MNRAS.364..217F,2005ApJS..159..242G}. In external galaxies, the situation is different. Because their disks can be observed with little line-of-sight extinction, the SNRs are typically identified through narrow-band optical imaging \citep{2015MNRAS.446..943V}.

Advances in X-ray astronomy have opened new prospects for detecting and studying SNRs. 
Until recently, however, ROSAT provided the only imaging all-sky soft X-ray survey, discovering several remnants (e.g. RX~J0852.0$-$4622, \citealt{1998Natur.396..141A} and references therein) but limited by modest sensitivity. Observatories such as XMM-Newton, Chandra, and XRISM study known SNRs in great detail, but have small fields of view (FoV) and cannot conduct systematic searches for new objects.
The eROSITA instrument, with its all-sky survey coverage and XMM-Newton-type sensitivity, offers a powerful tool for uncovering Galactic SNRs.
Besides, X-ray studies of diffuse emission provide information on electron temperatures, ionization states, and chemical abundances (for thermal emission); non-thermal X-rays trace synchrotron radiation from ultra-relativistic electrons in the magnetic fields, allowing to study cosmic ray acceleration. 

In this paper, we demonstrate the power of combining eROSITA X-ray observations with wide-field radio surveys to advance the study of Galactic supernova remnants in a limited region of the Circinus constellation. We identify new SNR candidates and revisit previously known remnants whose large-scale morphology and spectral properties were poorly constrained. 
In particular, we announce the discovery of two new SNR candidates,  designated as \GG\, and \blob\, and provide new information on SNRs MSH 15-52 and G320.6-1.6.
This pilot study shows that systematic, multiwavelength analyses even over limited sky regions can contribute to refining the census and characterization of Galactic SNRs. Extending this approach to the full sky (Becker et al., in prep) has the potential to provide a more complete and physically representative view of the Galactic SNR population.

\section{Identification and confirmation of SNRs}

The identification of an SNR relies on a combination multiwavelength diagnostics including morphological and spectral properties. 
In the radio band, SNRs typically show shell-like or centrally filled morphologies and non-thermal synchrotron emission, characterized by a negative spectral index \citep{2025JApA...46...14G,2008ARA&A..46...89R}. This synchrotron emission traces relativistic particles accelerated at supernova-driven shocks \citep{2015A&ARv..23....3D}.

The radio spectral index is useful for separating synchrotron-dominated remnants from thermal H\,\textsc{ii} regions. With the convention \(S_\nu \propto \nu^\alpha\), values of \(\alpha \lesssim -0.2\) generally indicate non-thermal emission, whereas H\,\textsc{ii} regions typically have flatter spectra, with \(\alpha \gtrsim -0.1\) \citep{2025A&A...693A.247A}. However, caution is required, since some SNRs show relatively flat radio spectra (e.g. \citealt{2008MNRAS.386.1411B}). Radio polarization can provide additional evidence for synchrotron emission, but does not by itself confirm an SNR classification \citep{2025A&A...693A.247A}.

Optical spectroscopy provides one of the most robust classification criteria. 
Shock-ionized gas in SNRs enhances forbidden line emission relative to recombination lines, leading to elevated line ratios such as $[\mathrm{S\,II}]/\mathrm{H}\alpha \gtrsim 0.4$, a threshold commonly used for distinguish SNRs from H\,\textsc{ii} regions \citep{1973ApJ...180..725M, 1997ApJS..108..261B}. 
Additional diagnostic lines include [O\,\textsc{i}], [O\,\textsc{ii}], and [O\,\textsc{iii}] relative to H$\alpha$ and H$\beta$ \citep{1985ApJ...292...29F, 2020MNRAS.491..889K}. However, for Galactic SNRs this criterion is often difficult to apply, as optical emission is detected for only $\sim 32\%$ of the currently known remnants \citep{2025JApA...46...14G}.

An additional discriminant between SNRs and H\,\textsc{ii} regions is the mid-infrared (MIR) to radio continuum flux ratio, which differs systematically between thermal and non-thermal sources \citep{2025A&A...693A.247A}.

In the X-ray band, SNRs typically exhibit emission from shock-heated plasma, often in non-equilibrium ionization (NEI) conditions, with typical temperatures of $kT \sim (0.2$--$5)$~keV \citep{2012A&ARv..20...49V}. Enhanced abundances of heavy elements such as O, Ne, Mg, Si, S, or Fe provide evidence for supernova ejecta and can further constrain the explosion type.

A secure confirmation generally requires consistent multiwavelength evidence, such as the coexistence of non-thermal radio emission, shock-excited optical spectra, and thermal X-ray emission from hot plasma. The detection of an associated compact object (neutron star or black hole) also strongly supports an SNR classification. Interactions with molecular clouds, traced for example by OH (1720~MHz) maser emission, provide additional confirmation of shock activity \citep{1996AJ....111.1651F}. 

Here we adopt a multiwavelength classification scheme for the sources discussed in this paper. The scheme is summarized in Tables~\ref{tab:snr_diagnostics} and~\ref{tab:snr_classification}. In all cases, we assume that the object has a morphology and angular size broadly consistent with an SNR; sources that clearly do not satisfy this basic requirement are not considered SNR candidates. We distinguish between decisive diagnostics, which can by themselves provide strong evidence for an SNR origin, and suggestive indicators, which support an SNR interpretation but are not sufficient for confirmation on their own. 

The diagnostics listed in Table~\ref{tab:snr_diagnostics} are not intended to be exhaustive, but represent the criteria most relevant for the radio, optical, infrared, and X-ray data used in this work.
We classify an object as a \emph{confirmed SNR} when at least one decisive diagnostic supports the SNR interpretation. Objects without a decisive diagnostic are classified as \emph{promising SNR candidates} when at least two independent suggestive indicators consistently support an SNR interpretation. Sources are classified as \emph{possible SNR candidates} when only one suggestive indicator is present, and the available data are insufficient to exclude common contaminants such as H\,\textsc{ii} regions, stellar-wind bubbles, or unrelated diffuse background emission.

\begin{table*}[t]
\centering
\caption{Multiwavelength diagnostic criteria used for the SNR classification in this work.}
\label{tab:snr_diagnostics}
\small
\begin{tabular}{p{0.4\linewidth}  p{0.55\linewidth}}
\toprule
\textbf{Decisive diagnostics} &
\textbf{Suggestive indicators} \\
\midrule
\begin{itemize}
    \item Optical shock-excited line ratios, e.g. [S\,\textsc{ii}]/H$\alpha \gtrsim 0.4$
    \item Measured non-thermal radio spectral index ($\sim -0.5$)
    \item Secure association with a compact object (neutron star or black hole)
    \item Expansion of a shell or filaments consistent with an SNR blast wave
\end{itemize}
&
\begin{itemize}
    \item Thermal X-ray emission consistent with shock-heated plasma
    \item Enhanced metal abundances indicating ejecta
    \item Filaments in optical forbidden lines 
    \item Radio polarization without a measured spectral index
    \item Absence of an obvious H\,\textsc{ii}-region counterpart
    \item MIR-to-radio flux ratio consistent with non-thermal emission
\end{itemize}
\\
\bottomrule
\end{tabular}
\end{table*}

\begin{table*}[t]
\centering
\caption{Working classification scheme used for SNR candidates in this paper. It is assumed that all candidates have a morphology and size consistent with an SNR.}
\label{tab:snr_classification}
\small
\begin{tabular}{p{0.28\linewidth} p{0.64\linewidth}}
\toprule
\textbf{Class} & \textbf{Adopted criterion} \\
\midrule

Confirmed SNR &
The source has at least one decisive diagnostic of an SNR origin.  \\

Promising SNR candidate &
The source lacks a decisive diagnostic, but at least two independent indicators consistently support an SNR interpretation. Such sources are considered strong targets for follow-up spectroscopy or deeper X-ray/radio observations. \\

Possible SNR candidate &
The source shows only one suggestive property.  \\

\bottomrule
\end{tabular}
\end{table*}

\section{Observations and data reduction}

In this work we used eROSITA data from the four completed All-Sky Surveys (abbreviated as eRASS:4) in the c020 processing version \citep{2024A&A...682A..34M}.
eROSITA (extended Röntgen Survey Imaging Telescope Array; \citealt{Predehl2021} is a highly sensitive X-ray instrument onboard the  German-Russian SRG (Spectrum Röntgen-Gamma) satellite \citep{Sunyaev2021}, launched in July 2019.
As of the time of writing (spring 2026) it has performed four and one third X-ray surveys of the sky in the soft X-ray band (\(\sim 0.2\text{--}8.0\) keV; \citealt{Brunner2022}), carried out between December 2019 and February 2022. It has seven telescope modules (TM1-TM7), each equipped with its own CCD camera, which provide a wide field of view (FoV) of $\sim 1^{\circ}$, and an angular resolution of 
26'' in survey mode (FoV averaged, \citealt{Predehl2021}). Given its all-sky survey coverage and its XMM-Newton type sensitivity, eROSITA turns out to be the perfect instrument to study extended sources having a low surface brightness such as SNRs.

The average on-source exposure time in the region of interest is approximately 1650\,s. After applying vignetting corrections, the effective exposure times are as follows: \(\sim\)830 s in the R band (0.2–0.7 keV), \(\sim\)850 s in the G band (0.7–1.1 keV), and \(\sim 690\)\,s in the B band (1.1–8.0 keV).
For data processing, we used eSASS (eROSITA Standard Analysis Software; \citealt{Brunner2022}), version 211214 (released in December 2021)\footnote{\url{https://erosita.mpe.mpg.de}}. The eSASS pipeline segments X-ray data into 4700 overlapping sky tiles, each covering $3.6^{\circ} \times 3.6^{\circ}$. These tiles are assigned six-digit identifiers, with the first three digits corresponding to right ascension (RA) and the last three to declination (Dec), indicating the sky tile centre in degrees.
For this study, we analysed data from eRASS sky tiles 224153, 226150, 230147, 231150, 231153, 237150, and 237153.

For the imaging analysis, we created images with the eSASS task \texttt{evtool} using a rebinning factor of \texttt{400}. In \texttt{evtool}, this factor refers to the virtual pixels which have a size of \(0.05''\) (not to be confused with the physical eROSITA CCD pixel size of \(9.6''\)). Therefore, \texttt{rebin=400} corresponds to a final image pixel size of \(400 \times 0.05'' = 20''.\)
This choice was motivated by the angular resolution of eROSITA in survey mode, which is \(\sim 26''\) on average over the field of view \citep{Predehl2021}. The adopted binning therefore reduces small-scale statistical noise while keeping enough spatial detail to preserve the features of the diffuse emission.

\begin{figure*}[!htbp]
    \centering
    \includegraphics[width=\linewidth]{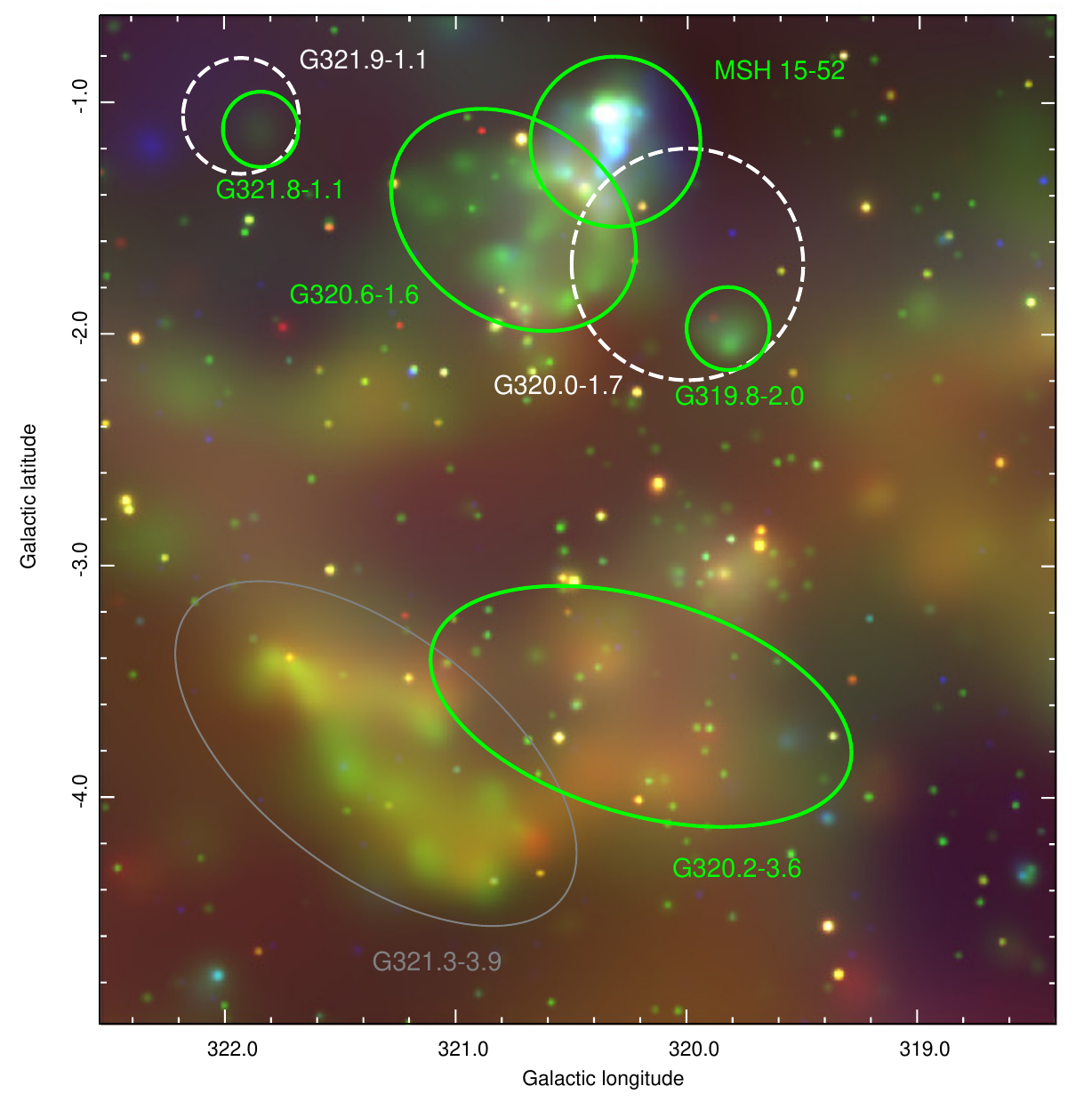}
    \caption{The RGB representation of the eRASS:4 image of the Circinus field in the energy bands R: 0.2-0.7 keV, G: 0.7-1.1 keV, B: 1.1-8.0 keV. 
    The image is corrected for exposure and vignetting and adaptively smoothed with a minimum significance of 3.  
    The same ASINH colour scaling is applied to all three bands, with displayed intensity limits of \(0\) and \(6\times10^{-6}\ \mathrm{cts\,cm^{-2}\,s^{-1}\,keV^{-1}}\). 
    }
    \label{fig:all_smooth_corr}
\end{figure*}

Fig.~\ref{fig:all_smooth_corr} shows an RGB image of the Circinus region using data from all seven telescope modules (TM1–7). To enhance the visibility of diffuse X-ray emission, we applied an adaptive kernel smoothing algorithm following \cite{Ebeling2006}. Specifically, we used the \texttt{csmooth} task from the CIAO package \citep{2006SPIE.6270E..1VF}, version 4.14, with a \texttt{sigmin} value of 3, while keeping all other parameters at their default settings.

A key goal of our image processing is to place all three energy bands (R/G/B) on a common physical scale so that a single set of intensity cuts can be applied. This ensures a neutral, physically meaningful colour representation to have a more objective way of selecting suitable RGB colour scale ranges. To achieve this, we divided the images in each band by the vignetting-corrected exposure map and normalized them by the different effective collecting areas of eROSITA across energy. The normalized image in band \(i\) is thus
\[
I_i^\ast(x,y) = \frac{I_i(x,y)}{E_i(x,y)\, C_i},
\]
where \(I_i\) is the raw counts image, \(E_i\) the exposure map, and
\[
    C_i = \int_{E_{i, low}}^{E_{i, high}} A_{\text{eff}} (E) \, dE,
\]
is the effective area (\(A_{\text{eff}}\)) integrated over the photon energy for band \(i\).
The resulting normalization factors are $C_R = 221.2$ cm$^{2}$ keV for the R band (0.2–0.7 keV), $C_G = 438.6$ cm$^{2}$ keV for the G band (0.7–1.1 keV), and $C_B = 2180.5$ cm$^{2}$ keV for the B band (1.1–8.0 keV).
With these operations, the fluxes in the R, G, and B bands become directly comparable, allowing us to apply identical cut values of \((0-6)\times10^{-6}~\mathrm{cts\,cm^{-2}\,s^{-1}\,keV^{-1}}\) in Fig.~\ref{fig:all_smooth_corr}. This procedure provides a reproducible and physically robust RGB representation that can be adopted in future studies.

\section{Spectral analysis method}

\label{sec:method}

X-ray spectra of SNRs are commonly described using a small set of standard models that represent the main emission processes. Thermal emission is modelled either as plasma in collisional ionization equilibrium (CIE) or as nonequilibrium ionization (NEI) plasma heated by shocks, while non-thermal emission is usually represented by a power-law.

The \texttt{apec} model describes an optically thin thermal plasma in or close to ionization equilibrium, characterized by a single temperature. In SNR studies, it is typically used to model older or denser shocked material, such as swept-up interstellar or circumstellar gas, where the ionization state has had time to adjust to the local temperature. This component produces emission lines from abundant elements together with a thermal continuum.

In contrast, young and middle-aged SNRs often show NEI conditions, where ionization lags behind rapid heating at the shock. The \texttt{nei} model represents a uniformly shocked plasma with a single temperature and ionization timescale. A more realistic description of shocked gas is provided by the \texttt{pshock}\footnote{{\bf P}lane-parallel Shocked plasma} (or \texttt{vpshock}\footnote{{\bf V}ariable-abundance Plane-parallel Shocked plasma.  Variable indicates that the elemental abundances (e.g., O, Ne, Mg, Fe) can be varied individually as free parameters, rather than being fixed to solar abundances.}) model, which accounts for a range of ionization timescales behind a plane-parallel shock and is commonly used for recently shocked material \citep{2001ApJ...548..820B}.

Non-thermal X-ray emission in SNRs, produced mainly by synchrotron radiation from relativistic electrons, is usually modelled with a power-law component. Such emission dominates the hard X-ray spectra of young shell remnants and pulsar wind nebulae. In practice, SNR spectra are often fitted with composite models that combine one or more thermal components with a power-law to account for both thermal plasma emission and non-thermal processes.

In all cases, the source emission models described above are modified by interstellar absorption. We model this absorption using the \texttt{TBabs} model \citep{2000ApJ...542..914W}, which accounts for photoelectric absorption by neutral gas along the line of sight. Unless stated otherwise, \texttt{TBabs} is applied multiplicatively to all thermal and non-thermal spectral components, with the  hydrogen column density $N_{\rm H}$ treated as a free parameter in the fits.

We conducted the X-ray spectral analysis of the eROSITA survey data using \texttt{PyXSPEC}, the Python implementation of \texttt{XSPEC} \citep{Arnaud1996}. For data extraction, we only used eROSITA telescope modules (TMs) 1-4 and 6 due to the light leak in the TM5 and TM7 \citep{Predehl2021}, which makes their spectral calibration uncertain, albeit their use for imaging seems not to be problematic. 
For spectral fitting, we employed the Cash statistic \citep{Cash1979} as implemented in \texttt{XSPEC}, reporting errors at the $1 \sigma$ confidence level. 

To assess the goodness-of-fit we performed Monte Carlo goodness tests using the implementation available in \texttt{XSPEC}. This method estimates the fraction of simulated spectra, generated from the best-fit model, that yield a fit statistic lower than that obtained for the observed data. Values close to 50\% indicate that the model provides an adequate statistical description of the data, whereas values approaching 0\% or 100\% suggest systematic discrepancies between the model and the observations.

For model comparison, we additionally employed the Akaike Information Criterion (AIC; \citealt{1974ITAC...19..716A}), defined as $\mathrm{AIC} = 2k - 2\ln\mathcal{L}$, where $k$ is the number of free parameters and $\mathcal{L}$ the maximum likelihood. The AIC penalizes models with additional free parameters and thus balances goodness-of-fit against model complexity. Differences in AIC values ($\Delta \mathrm{AIC}$) were used to assess the relative support for competing models, with lower AIC values indicating a statistically preferred model.

To minimize contamination from unrelated point sources in our spectral analysis, we excluded point sources within the target extended source boundaries with a logarithmic detection likelihood of DETLIKE $= -\ln({\rm P}) \ge 30$.
Although this threshold may appear stringent, it is necessary to avoid the removal of numerous false point sources produced by the detection algorithm, which tends to interpret diffuse emission as multiple individual point sources. This effect is particularly pronounced in SNRs, where bright extended emission often leads to over-densities of falsely identified point sources \citep{2024A&A...682A..34M}.

Point sources with detection likelihoods below this threshold ($\mathrm{DET\_LIKE} < 30$) were assumed to have a negligible contribution compared to the overall SNR emission. We estimated the systematic uncertainty introduced by the inclusion of such faint sources in the diffuse emission analysis to be approximately $2\%$ (see Appendix~\ref{sec:appendix}). All identified point sources were masked out using circular regions with a radius of $120^{\prime\prime}$, based on the instrument’s point spread function.

To ensure an accurate treatment of the background, we did not simply subtract a background spectrum. Instead, for each source we defined source and background extraction regions and fitted them simultaneously, explicitly modelling the background emission. Our baseline background model follows the approach described by \citet{2023A&A...676A...3Y} and includes local hot bubble,
Milky Way's circum-galactic medium, Galactic corona, cosmic X-ray background, and instrumental background. However, because our fields lie close to the Galactic plane, we included an additional component to account for the Galactic Ridge X-ray Emission (GRXE), which is not part of the original model. This component is particularly important for some objects, including G320.6–1.6, where the background spectrum cannot be satisfactorily reproduced without it. The impact of the GRXE component is discussed in detail in Sect.~\ref{sec:spectrum_G320.6_GRXE}.

\section{Results}

\subsection{New SNR candidate \GG}

\begin{figure}[htb!]
    \centering
    \includegraphics[width=\linewidth]{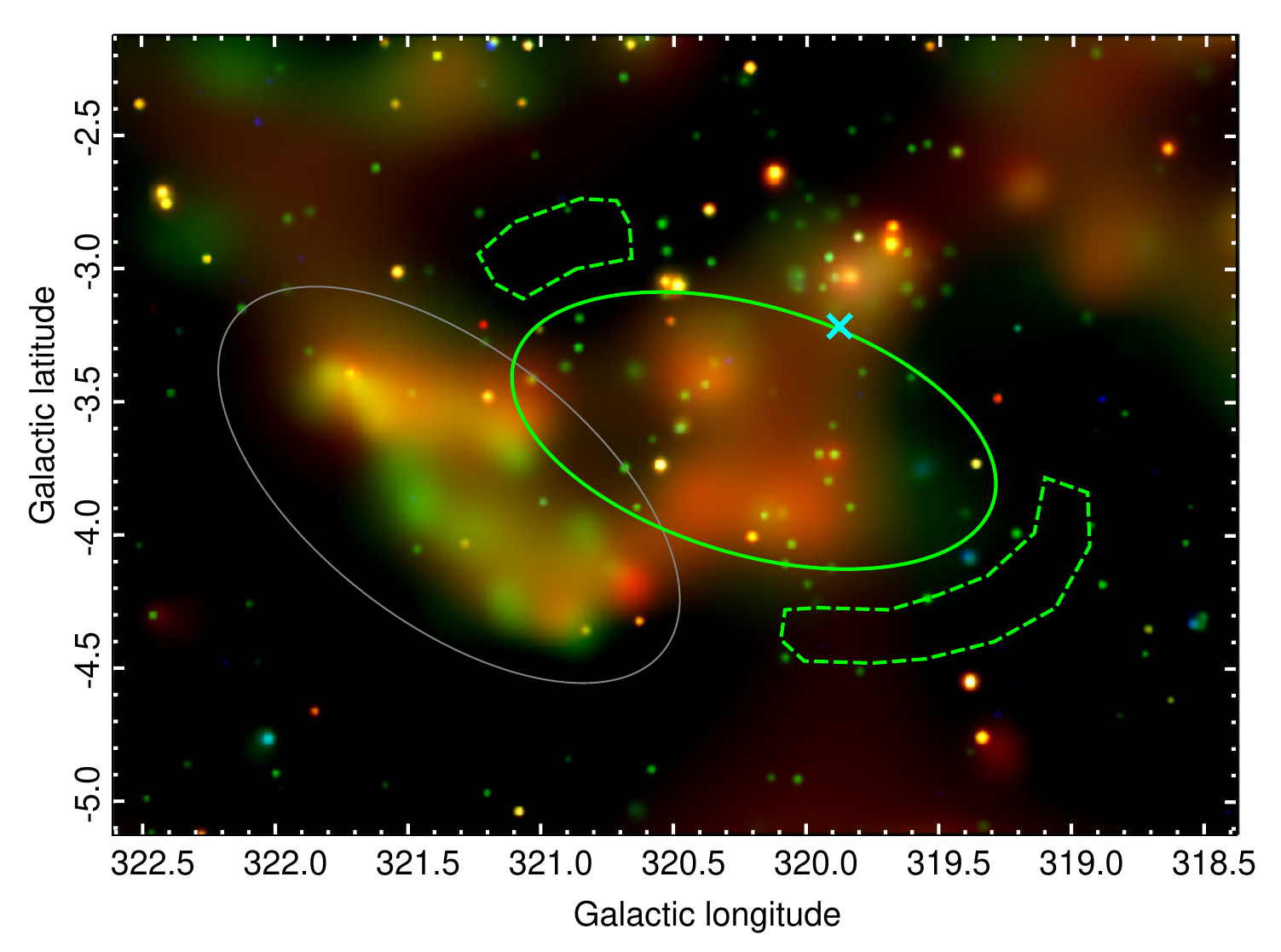}
    \caption{
    Adaptively smoothed eRASS:4 image showing the extraction regions used for the spectral analysis. The grey contour outlines the known SNR G321.3-3.9, the solid green contour marks the newly identified SNR candidate \GG, and the green dashed regions indicate the background areas used for the spectral analysis of \GG. The background regions were defined along the observatory’s scan direction. The overlapping region with G321.3-3.9 was excluded from the \GG \; source region to avoid contamination.
    The cyan cross marks the position of the PSR J1519-6106.
    }
    \label{fig:G320}
\end{figure}

We have identified a new SNR candidate, designated \GG.
In the eRASS:4 data, this candidate presents a diffuse structure with dimensions of approximately  $ 0.9 \times 0.5 $ degrees as shown by the yellow ellipse in Fig.~\ref{fig:G320}).
\GG\, is located to the west of the known, recently confirmed SNR G321.3-3.9 \citep{2024Mantovanini,2024Fesen,2014PASA...31...42G,1997MNRAS.287..722D}.

\begin{figure*}
    \centering
    \includegraphics[width=\linewidth]{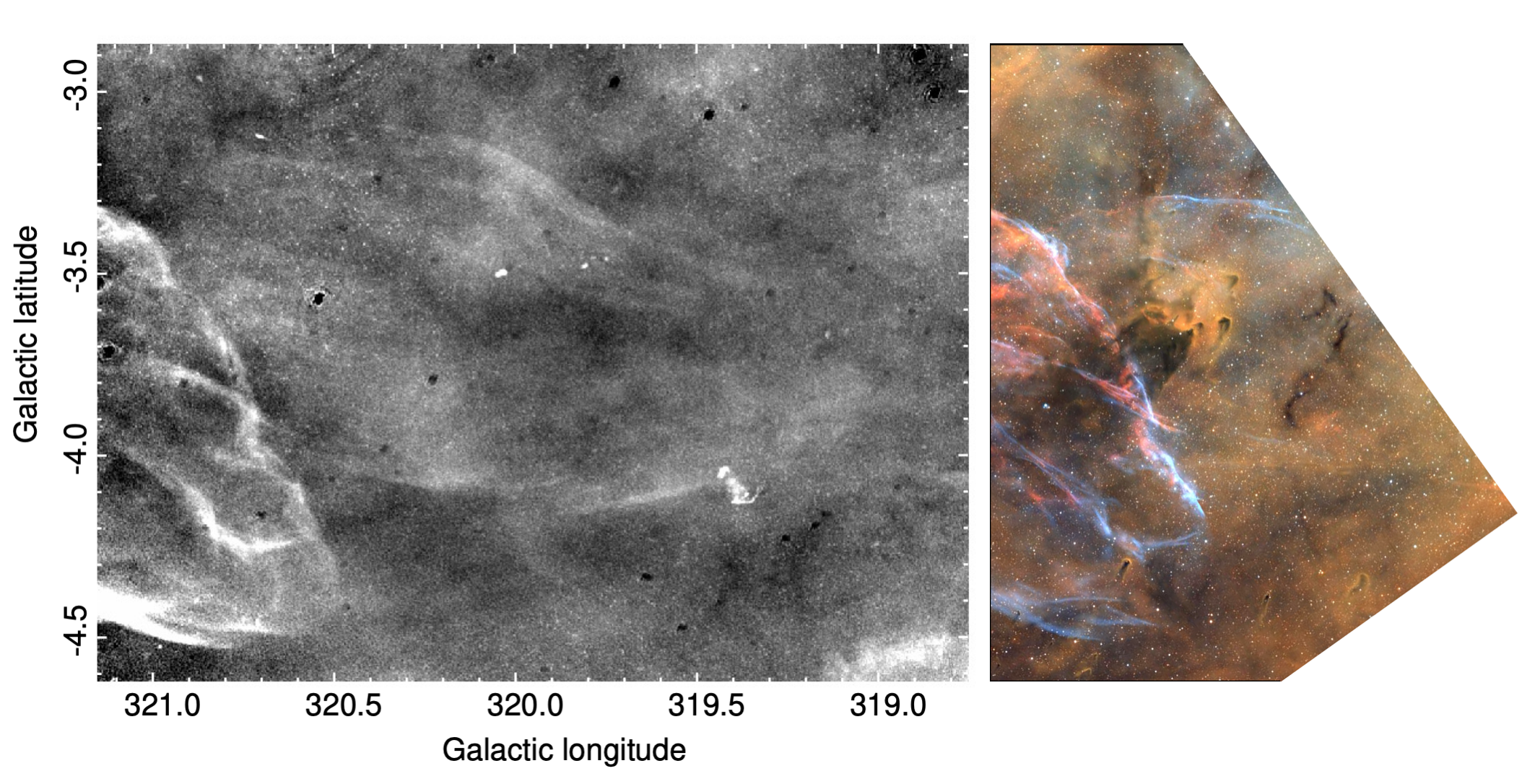}
    \caption{Left: \GG\ as seen in EMU at 943 MHz. We removed the radio point sources to the greatest extent to better show the radio shell. The remnant partly visible to the east is SNR G321.3-3.9. Right: Optical colour-composite image (courtesy of \citealt{2024Fesen}) covering part of the field shown in the radio image on the left at the same scale. 
    Optical filaments of G320.2$-$3.6 are visible at the upper west, mostly in blue, originating from [O III] emission.}
    \label{fig:G320-EMU}
\end{figure*}

In addition to the eROSITA detection, \GG  \; has been detected in radio wavelengths through the Evolutionary Map of the Universe (EMU) survey at 943 MHz (see the left panel of Fig.~\ref{fig:G320-EMU}, \citealt{2021PASA...38...46N}). 
The candidate remnant is also visible in narrow-band optical data (right panel of Fig.~\ref{fig:G320-EMU}; \citealt{2024Fesen}), and was described by \citet{2024Fesen} as a possible ``blowout feature'' of G321.3$-$3.9.
However, given its shell-like morphology, and its size, which is similar to that of G321.3$-$3.9, we consider \GG\ to be a separate object.

In the infrared, we examined the WISE data and the WISE Catalogue of Galactic HII Regions V2.3\footnote{\url{https://astro.phys.wvu.edu/wise/}} \citep{2014ApJS..212....1A, 2015ApJS..221...26A} to assess the presence of H\,II regions. We find no evidence for H\,II regions at this location, which is a supporting diagnostic for a possible SNR interpretation.

In optical wavelengths, \GG \; displays emission in H$_{\alpha}$, [O III], and [S II] \citep{2024Fesen}. The optical observations indicate that the candidate SNR's shell is currently in a radiative phase, consistent with a late-stage, cooling evolved SNR. 
The overall morphology of \GG \; is elongated. The radio emission traces the optical filaments.

\begin{figure}
    \centering
    \includegraphics[width=\columnwidth]{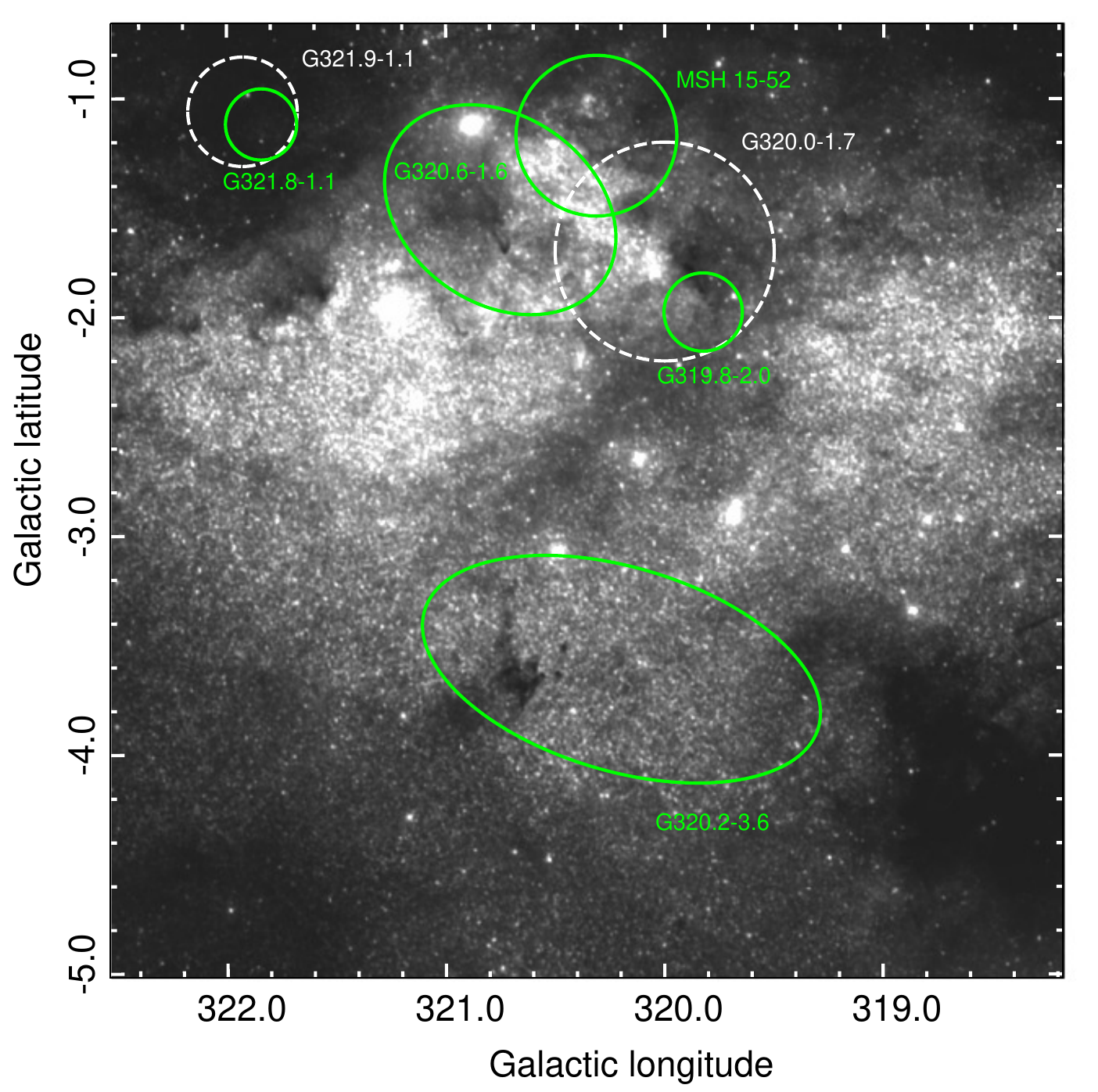}
    \caption{Optical image of the same region as Fig.~\ref{fig:all_smooth_corr} from the Digitized Sky Survey (DSS2 red). A comparison with the X-ray image shows the effect of X-ray shadowing, where dense interstellar material, visible as dark clouds in the optical image, absorbs soft X-ray emission and produces corresponding deficits in the X-ray map. 
    }
    \label{fig:dss2}
\end{figure}

The X-ray emission detected in eRASS:4 does not follow the shell structure observed in radio and optical wavelengths, but instead fills the interior (see Fig.~\ref{fig:G320}). 
However, the morphology of this emission, is affected by absorption.
A comparison between the eROSITA image and the Digitized Sky Survey (DSS2) optical data shows that several depressions in the X-ray emission coincide with regions of high optical extinction (Fig.~\ref{fig:dss2}). This shadowing effect, produced by dust/molecular clouds, shapes much of the apparent X-ray morphology, including the  ``hole'' between G321.3-3.9 and \GG. Consequently, the observed X-ray morphology might be caused by variations in foreground absorption rather than by intrinsic emission from the candidate SNR.

Given this, it remains unclear whether the detected X-ray emission originates from the candidate remnant or from unrelated foreground or background components. We therefore treat the X-ray detection as suggestive, but not sufficient to confirm an association.

The X-ray emission at the location of \GG \; appears to be softer compared to  G321.3$-$3.9 (see Fig.~\ref{fig:G320}). To quantify this apparent colour difference in the RGB image, we compared the number of counts in the red and green energy bands used for the composite. For G321.3$-$3.9 we measure approximately $1.1\times10^4$ and $1.8\times10^4$ counts in the R and G bands, respectively, whereas \GG\ contains $\sim1.0\times10^4$ and $\sim1.4\times10^4$ counts in the same bands. 
The B band is not considered here because both sources show only background-level counts in this energy range. The colour difference between the two objects is therefore driven by the relative contribution of the green band: G321.3$-$3.9 has a stronger green-band component compared to \GG. This difference in the spectral distribution of counts is consistent with, but does not uniquely support the interpretation that \GG\ and G321.3$-$3.9 represent distinct physical systems rather than components of a single extended structure.

A known pulsar, PSR~J1519$-$6106 (RA $=229.8992^\circ$, Dec $=-61.1152^\circ$), is located near the northern edge of \GG. 
Its characteristic spin-down age is $\sim4$~Myr, which is more than an order of magnitude larger than 
the mean lifetime of a supernova remnant ($6 \times 10^4$ yrs \citealt{1994ApJ...437..781F}).
Assuming the spin-down age provides a realistic estimate of the pulsar's true age, such a discrepancy makes a physical association between the pulsar and \GG\ unlikely. 
Cross-correlating the pulsar's position with the eROSITA point-source catalogue generated by the eSASS pipeline (\citealt{2024A&A...682A..34M,Brunner2022}, 1B detection process in the 0.2-2.3 keV band), we find a low-significant ($\sim 4\sigma$) X-ray point source at $=229.8909^\circ$, Dec $=-61.1009^\circ$. Given its angular separation of $54''$ from the pulsar position and eROSITA's positional point source accuracy of about $6\pm 2$ arcsec (95\% error) for a $\sim 4\sigma$ source \citep{2024A&A...682A..34M} we consider this low-significant X-ray source to be unrelated with the pulsar. 

\subsubsection{Spectral analysis}

\begin{figure*}
  \centering

  \begin{subfigure}{0.45\textwidth}
    \includegraphics[width=\linewidth]{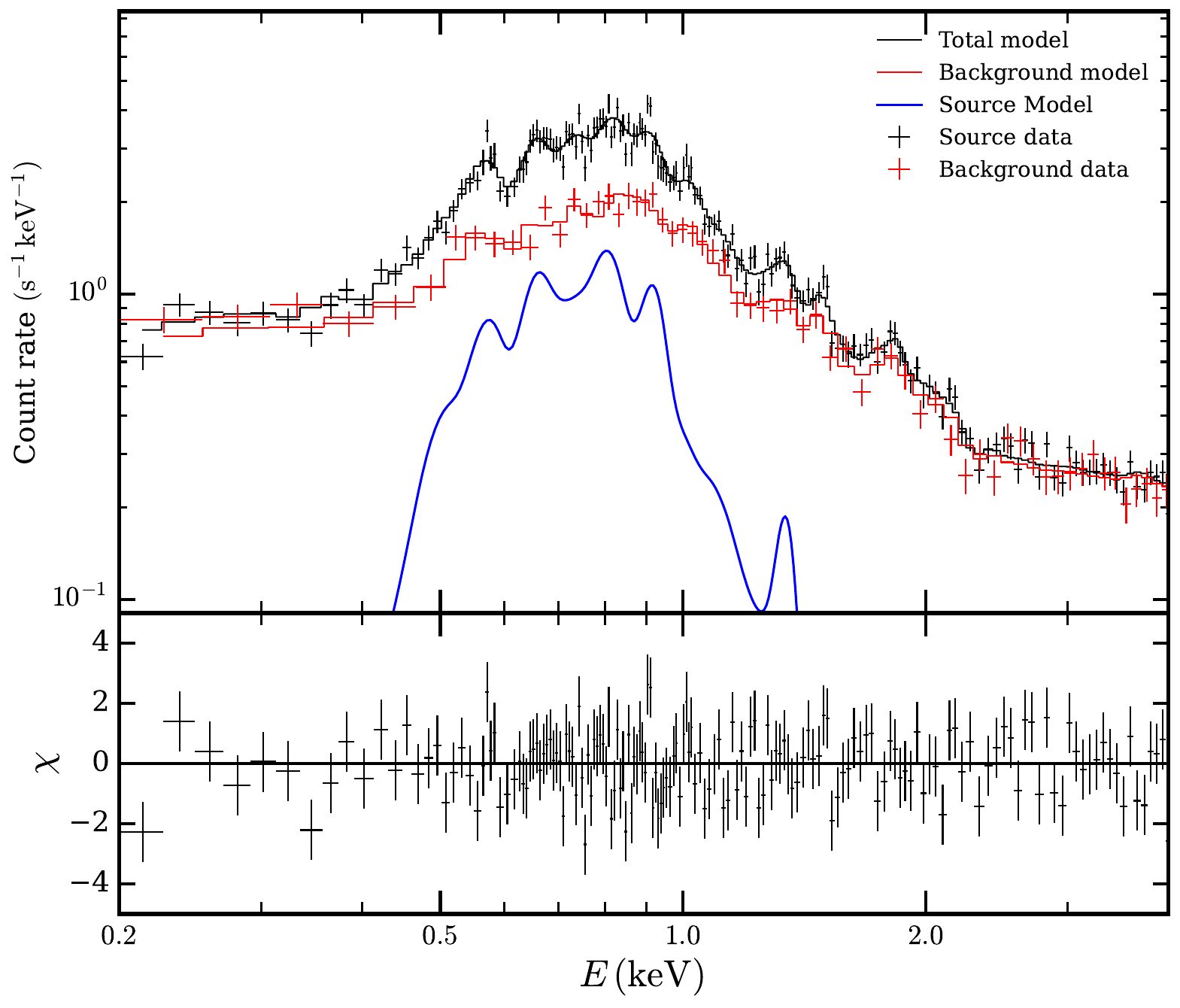}
    \caption{ Spectrum of \GG, fitted with a \texttt{vapec} thermal plasma model.}
    \label{fig:G320.0_spectrum}
  \end{subfigure}
  \hfill
  \begin{subfigure}{0.45\textwidth}
    \includegraphics[width=\linewidth]{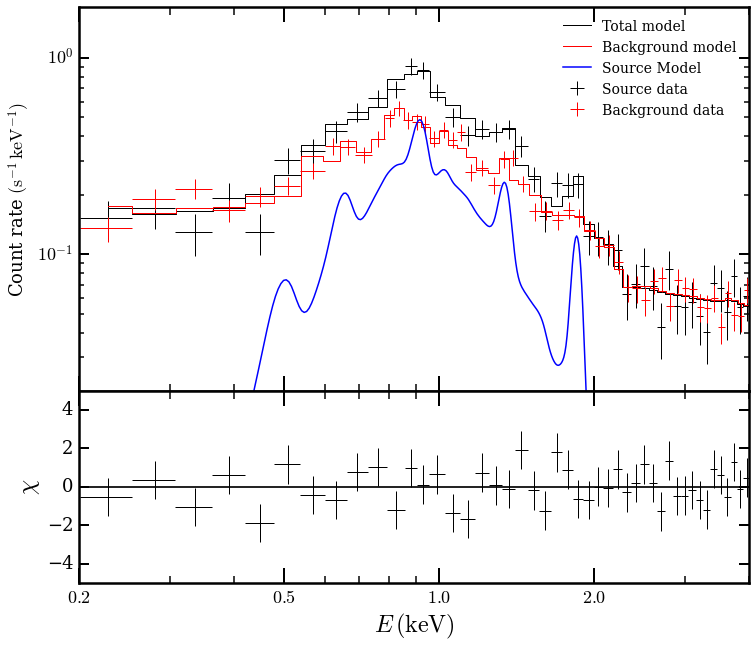}
    \caption{Spectrum of \blob, fitted with a \texttt{vapec} thermal plasma model.}
    \label{fig:blob_spectrum}
  \end{subfigure}

  \medskip

  \begin{subfigure}{0.45\textwidth}
    \includegraphics[width=\linewidth]{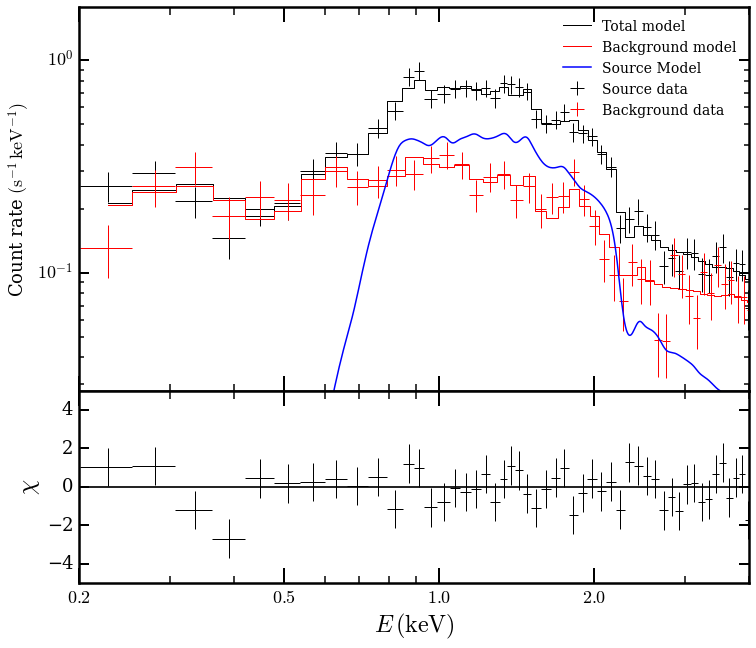}
    \caption{Spectrum of the central diffuse nebula in MSH 15-52, fitted with a combined (powerlaw+vapec) spectral model.}
  \end{subfigure}
  \hfill
  \begin{subfigure}{0.45\textwidth}
    \includegraphics[width=\linewidth]{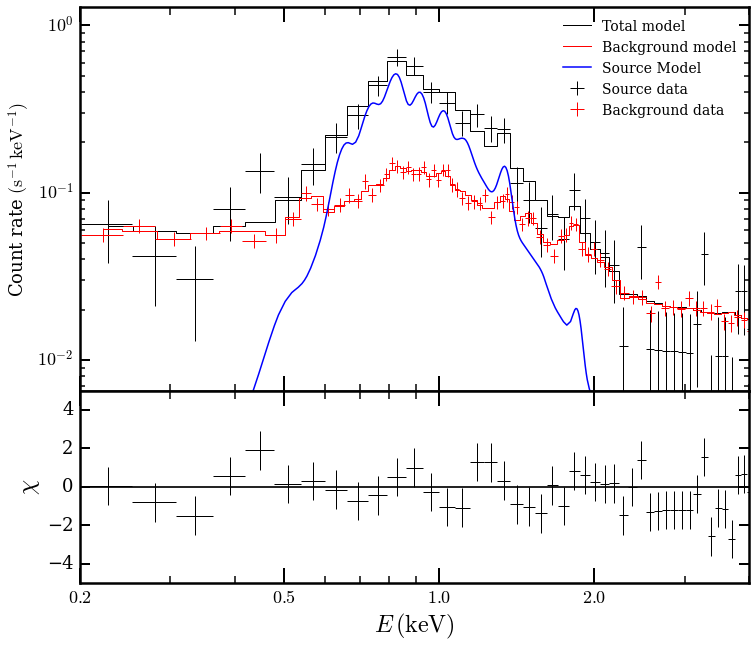}
    \caption{Spectrum of G320.5-1.8, fitted with a \texttt{apec} thermal plasma model.}
    \label{fig:G3205-spectrum}
  \end{subfigure}

  \medskip

  \begin{subfigure}{0.45\textwidth}
    \includegraphics[width=\linewidth]{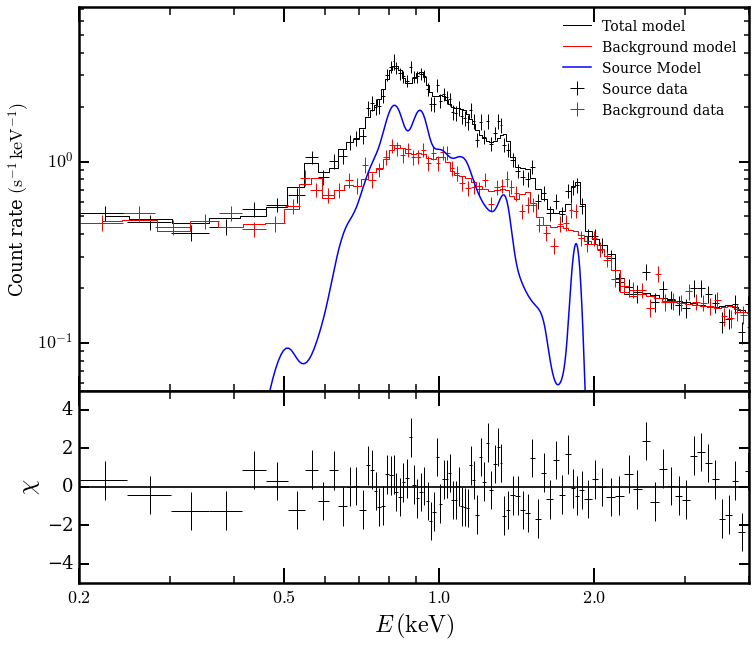}
    \caption{Spectrum of G320.6-1.6, fitted with a \texttt{vapec} thermal plasma model.}
    \label{fig:G3206_spectr}
  \end{subfigure}
  \hfill
  \begin{subfigure}{0.45\textwidth}
    \includegraphics[width=\linewidth]{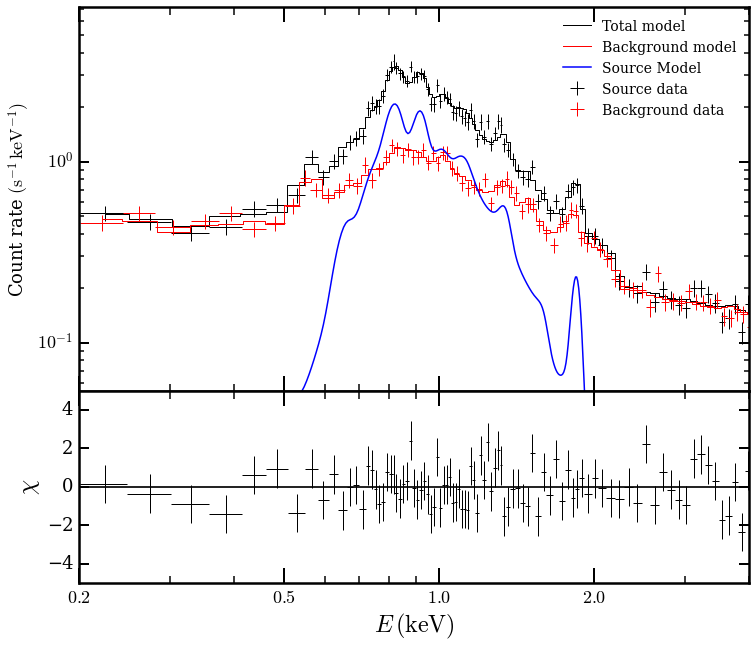}
    \caption{Spectrum of G320.6-1.6, fitted with a \texttt{vapec} thermal plasma model. With the GRXE component in the background model.}
  \end{subfigure}

  \caption{eROSITA X-ray spectra of the objects analysed in this work, with best-fit spectral models overlaid.}
\end{figure*}

We analysed the X-ray spectrum of \GG \; using the extraction regions indicated in Fig.~\ref{fig:G320}. 
The spectrum of this candidate is well described with both the \texttt{vpshock} and \texttt{apec}/\texttt{vapec} models (see Table~\ref{tab:G320} and Fig.~\ref{fig:G320.0_spectrum}).
While all models provide statistically acceptable fits, they yield different best-fit parameter values. In particular, the inferred absorbing column density varies between the models, with $N_H/10^{22}~\mbox{cm}^{-2}$ 
values of $0.13^{+0.09}_{-0.06}$ (vpshock), $0.56^{+0.10}_{-0.18}$ (apec), and $0.79^{+0.09}_{-0.11}$ (vapec). These differences indicate that the spectral parameters are not uniquely constrained by the current data, reflecting degeneracies between absorption, plasma temperature, and ionization state. As a result, the available data do not allow us to discriminate robustly between the considered thermal models. Therefore, the physical interpretation of this emission remains uncertain.

\begin{table}[]
\centering
\caption{\GG. Best fit values obtained with different models. For the \texttt{apec} model, all metal abundances are frozen at 1.
\label{tab:G320}}
\begin{tabular}{lccc}
\toprule
Model & vpshock & vapec & apec \\%
\midrule
$N_{\text{H}}$ (10$^{22}$ cm$^{-2}$) & $0.13_{-0.06}^{+0.09}$ &$0.79_{-0.11}^{+0.09}$  & $0.56_{-0.18}^{+0.10}$ \\[1ex]%
kT (keV) & $0.56_{-0.11}^{+0.08}$ & $0.15_{-0.01}^{+0.01}$ &  $0.22_{-0.02}^{+0.04}$ \\[1ex]%
O/O$_{\odot}$ &  $1.6_{-0.8}^{+1.1}$ & $0.4_{-0.2}^{+0.3}$  & - \\[1ex]%
Ne/Ne$_{\odot}$ &  $1.8_{-0.6}^{+1.0}$ & $0.5_{-0.2}^{+0.4}$& - \\[1ex]%
Mg/Mg$_{\odot}$ & $1.7_{-0.9}^{+1.1}$ &$1.5_{-0.9}^{+0.8}$  & -\\[1ex]%
Si/Si$_{\odot}$ & =1 & $0.1_{-0.1}^{+2.5}$ & -  \\[1ex]%
Fe/Fe$_{\odot}$ & =1 &  >5.1   &  -\\[1ex]%
Tau$_u$ (10$^{11}$ cm$^{-3}$ s)  & $6_{-3}^{+8}$ & - & - \\[1ex]%
Normalization & $0.0025_{-0.0005}^{+0.0010}$  & $0.7_{-0.4}^{+0.7}$  &  $0.04_{-0.03}^{+0.03}$  \\[1ex]%
\midrule
CStat & 1722 & 1711  &   1726 \\ %
d.o.f. & 1593 & 1592 & 1597 \\
AIC & 1776  &  1767  &  1772 \\%
\( \Delta\) AIC & 9 & 0 & 5 \\
goodness & 81\% & 80\% & 82\% \\
\bottomrule
\end{tabular}
\end{table}

For comparison, the absorbing column density derived for G321.3–3.9 lies in the range $(0.09$–$0.19)\times10^{22}$~cm$^{-2}$, which is consistent with the value obtained from the \texttt{vpshock} fit for \GG, but lower than that inferred from the \texttt{apec}/\texttt{vapec} models. This suggests that, depending on the adopted spectral model, \GG \; may be located at a distance comparable to that of G321.3–3.9 or potentially at a larger distance. However, given the model-dependent uncertainties in $N_{\rm H}$, this conclusion remains tentative.

The plasma temperature varies between the models from  $0.15_{-0.01}^{+0.01}$ keV (vapec) to $0.56_{-0.11}^{+0.08}$ keV (vpshock).
The latter value is similar to that obtained for G321.3-3.9 \citep{2024Mantovanini}, where different models yield \(kT = (0.55-0.68)~\mathrm{keV}\) (1$\sigma$ range). 

Note that the spectral parameters are not uniquely constrained; therefore, the physical interpretation remains uncertain.

\subsection{Candidate SNR \blob} 

During our inspection of the eROSITA data, we identified a compact region of diffuse X-ray emission located to the south of MSH 15-52 and G320.6-1.6.
This structure has an angular size of $\approx 10'$ and exhibits a centrally concentrated brightness peak at Galactic coordinates (l,b) = (319.8$^{\circ}$, -2.07$^{\circ}$) (see south-west corner of Fig.~\ref{fig:X+radio}).

\begin{figure}
    \centering
    \includegraphics[width=\columnwidth]{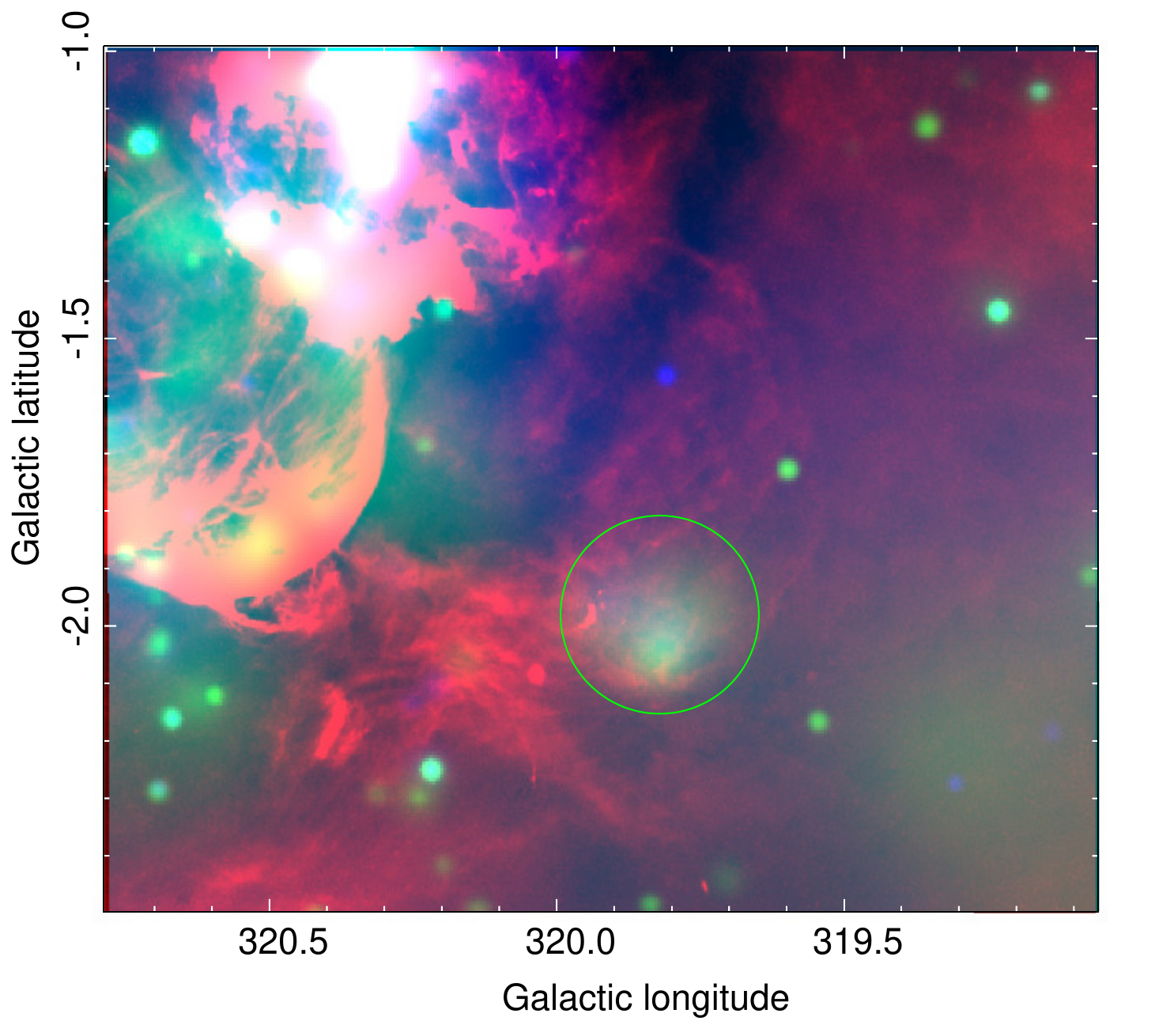}
    \caption{RGB image of \blob\ (indicated by the green circle) and G320.0$-$1.7. Red: EMU radio data with point sources removed. Green and blue: eRASS:4 X-ray data, adaptively smoothed.}
    \label{fig:blob}
\end{figure}

This X-ray source lies within the boundaries of the candidate SNR G320.0–1.7  (\citealt{2025ApJ...988...75B}, see in Fig.~\ref{fig:blob}). However, given that the extent of the detected X-ray emission is smaller than that of G320.0–1.7, the association between the two sources appears unlikely.

\blob was already noted in the ROSAT All-Sky Survey as a possible SNR candidate (cf. Prinz 2013). 
That work also reported extended infrared emission potentially associated with the X-ray feature, as well as tentative evidence for filamentary optical structures in its vicinity. 

We checked the infrared WISE data and the WISE Catalogue of Galactic HII Regions V2.3 \citep{2014ApJS..212....1A, 2015ApJS..221...26A} for the presence of H\,II regions. We find no evidence for H\,II regions at this location.

Our data allow us to re-examine this object with eROSITA. The morphology of this source can be seen in Fig.~\ref{fig:X+radio}. While the southern boundary appears sharp and well defined, the northern side exhibits a more gradual transition into the surrounding background. The emission displays an approximate symmetry along the east--west axis and is slightly elongated in the direction of Galactic longitude. 

To further investigate the physical nature of \blob, we performed spectral analysis using various plasma emission models. The spectrum is best described by thermal models; nonthermal simple power-law model failed to provide a good fit. 
Our initial modelling employed the vapec model, which produced a good fit to the data (see Fig.~\ref{fig:blob_spectrum} and Table~\ref{tab:blob}). We also tested the vpshock model; however, it did not yield any statistically significant improvement over vapec, and the ionization timescale parameter ($\tau_{\text{u}}$) remained unconstrained. Consequently, we adopted the vapec model for subsequent analysis.

\begin{table}[]
\centering
\caption{\blob. Best fit values obtained with different models with O, Ne, Mg, Si, and Fe free to vary. 
\label{tab:blob}}
\begin{tabular}{lc}
\toprule
Model &  vapec   \\%
\midrule
$N_{\text{H}}$ (10$^{22}$ cm$^{-2}$) &  $1.2_{-0.5}^{+0.2}$  \\[1ex]%
kT (keV) &  $0.20_{-0.03}^{+0.12}$ \\[1ex]%
O/O$_{\odot}$   &   $0.8_{-0.7}^{+5}$ \\[1ex]%
Ne/Ne$_{\odot}$ &   $0.6_{-0.4}^{+2}$  \\[1ex]%
Mg/Mg$_{\odot}$ &  $1.2_{-0.7}^{+3}$  \\[1ex]%
Si/Si$_{\odot}$  &   $> 3$  \\[1ex]%
Fe/Fe$_{\odot}$ &    $ < 0.5 $   \\[1ex]%
Normalization &   $0.06_{-0.04}^{+0.23}$  \\[1ex]%
\midrule
Statistic/d.o.f.  &    1670/1600  \\ %
goodness & 18\% \\
\bottomrule \\
\end{tabular}
\end{table}

The hydrogen column density ($N_\text{H}$) is not well constrained in our fits, likely due to limited photon statistics and degeneracies with other spectral parameters. However, we obtain a reasonably well-defined temperature estimate of approximately $kT \approx $ 0.2~keV. 
Elemental abundances are challenging to determine, with all elements exhibiting large uncertainties. While a simpler apec model was also tested, it resulted in a poor fit, with the abundance parameter remaining unconstrained. 

To investigate the nature of the X-ray emission in \blob, we examined the Milky Way molecular clouds from $^{12}$CO  (MML2017) catalogue \citep{2017ApJ...834...57M} for the presence of molecular clouds. While several clouds are located within the boundaries of G320.0$-$1.7, we find no clear indications of interaction. In particular, establishing such an interaction would require molecular-line observations to test for broadened or disturbed line profiles (e.g. \citealt{2000ApJ...545..874R} and references therein).
The detection of OH(1720 MHz) maser emission would provide evidence for an interaction with molecular gas \citep{1996AJ....111.1651F}. However, no such masers are reported within G320.0$-$1.7 in the database of circumstellar OH masers \citep{2015A&A...582A..68E}. 
Gamma-ray emission spatially coincident with dense gas could also indicate interaction \citep{2010ApJ...717..372C}, but no corresponding source is listed in the \textit{Fermi}-LAT Fourth Source Catalogue \citep{2022ApJS..260...53A}.
An interpretation as a pulsar wind nebula is also disfavoured, as the X-ray emission is better described by a thermal model, and no corresponding radio PWN is detected in the EMU data.
Deeper X-ray observations are needed to draw conclusions about the physical interpretation of this emission.

\subsection{G321.9-1.1}
\label{sec:snrpy_G321.9}

\begin{figure*}
    \centering
    \includegraphics[width=0.5\linewidth]{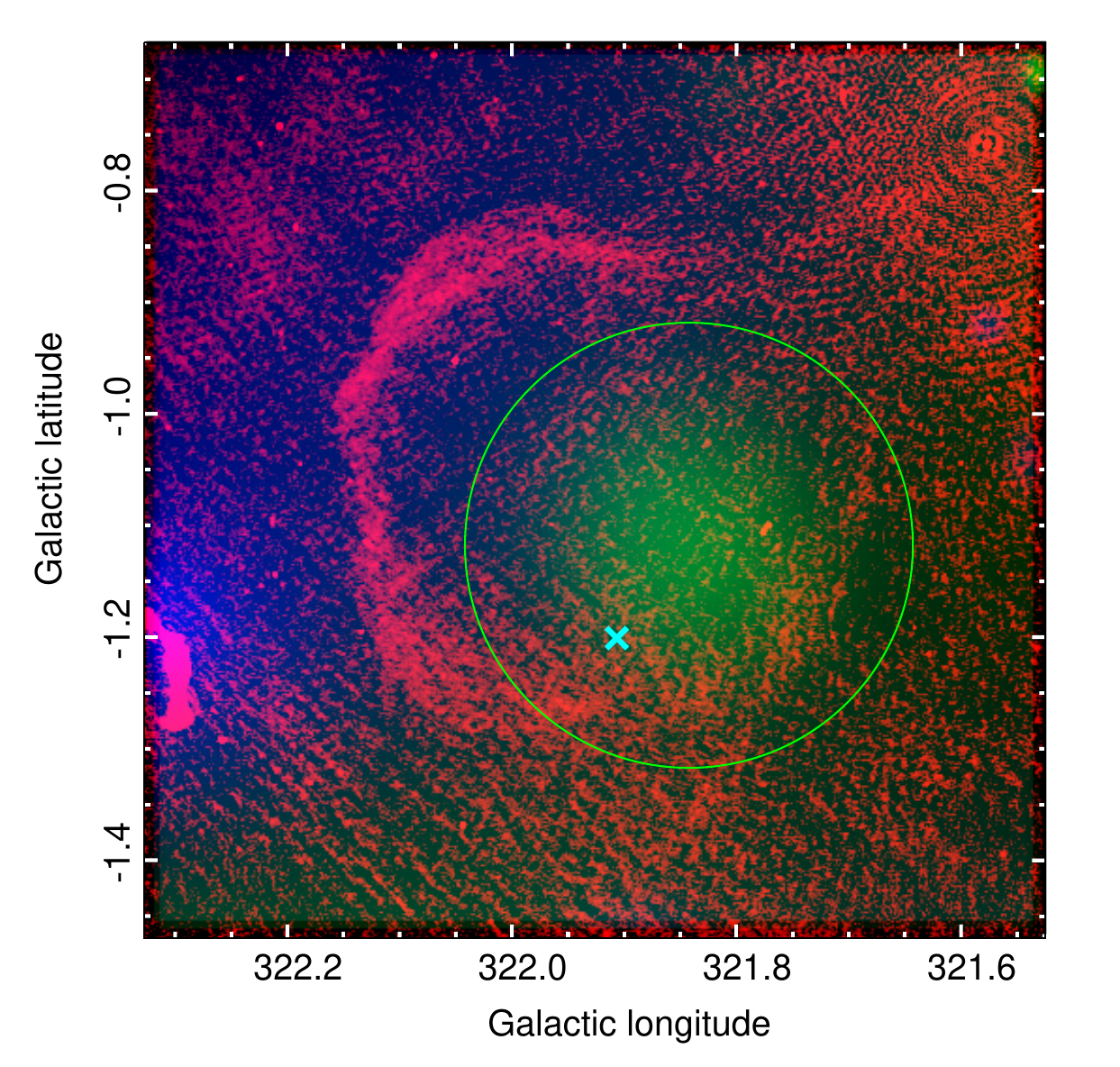}%
        \includegraphics[width=0.5\linewidth]{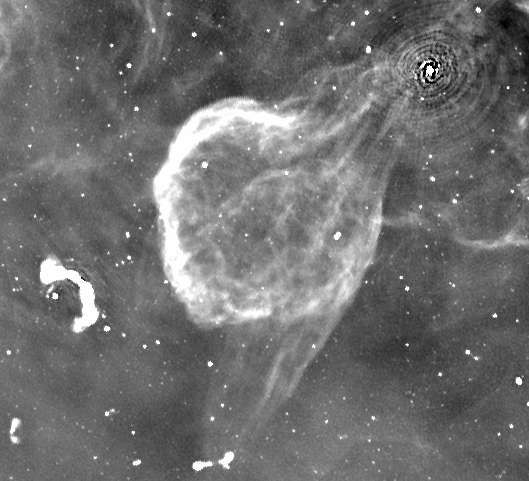}
    \caption{Left: composite RGB image of the G321.9-1.1 region. Red: MeerKAT 1.3 GHz (SMGPS); green: eRASS:4 data in the 0.7--1.1 keV band; blue: eRASS:4 data in the 1.1--8.0 keV band. The X-ray images are adaptively smoothed with a minimum significance of 3. The green circle indicates the region used for the spectral analysis, and the cyan cross marks the position of the pulsar PSR~J1524$-$5819. Right: EMU 943 MHz.}
    \label{fig:G321.9-1.1}
\end{figure*}

We detect diffuse X-ray emission with eROSITA that is spatially coincident with part of the radio SNR G321.9$-$1.1 (e.g. \citealt{2025JApA...46...14G}), suggesting a possible association (see left image in Fig.~\ref{fig:G321.9-1.1}).  This remnant has so far been detected only in the radio band (e.g. \citealt{1996A&AS..118..329W,2025A&A...693A.247A}). No radio spectral index measurement is available in the literature, but \citet{2025ApJ...988...75B} report a flux density of \(3.9 \pm 0.3\) Jy at 943 MHz.
The radio emission shows a roughly circular shell with a radius of \(\sim 15'\). Inspection of the EMU radio image in Fig.~\ref{fig:G321.9-1.1} suggests the presence of a slightly off-centre inner circular or filamentary shell, which we tentatively refer to as G321.8-1.1. It is not clear whether this structure is physically distinct from G321.9–1.1 — for example, representing a possible remnant within a remnant — or whether it is part of the same system. Similarly, the apparent association with the nearby bright radio source, MGPS~J152029$-$580747 \citep{2007MNRAS.382..382M} with a flux density of $\sim 3.6$ Jy\,beam$^{-1}$ at 843 MHz, to which several linear filaments appear to connect, remains uncertain. 

Remarkably, the diffuse emission detected by eROSITA appears to be spatially coincident with this putative secondary ``inner shell''. Although the significance of the extended eROSITA source in the G-band is only at the level of $\sim 3$–$4\sigma$ above the background, and the photon statistics are limited to $\sim 100$ source counts, it may represent a potentially interesting X-ray counterpart to a particular region of G321.9–1.1, though the significance of this correspondence requires further investigation.

Despite the limited photon statistics, we carried out a preliminary and qualitative spectral characterization of the X-ray emission using a simple \texttt{apec} model. The fit yields a hydrogen column density of \(N_{\mathrm H} = 1.7^{+0.5}_{-0.4} \times 10^{22}\ \mathrm{cm^{-2}}\) and a plasma temperature of \(kT = 0.47^{+0.13}_{-0.12}\ \mathrm{keV}\). Despite the limited data quality, these values are consistent with those typically observed in  supernova remnants ($kT \sim (0.2$--$5)$~keV, \citealt{2012A&ARv..20...49V}).

We searched the infrared WISE data and the WISE Catalogue of Galactic HII Regions V2.3 \citep{2014ApJS..212....1A, 2015ApJS..221...26A} for the presence of H\,II regions. One H\,II region, WISE G321.921$-$0.921, is located within the projected boundaries of G321.9$-$1.1 at the northern remnant edge. Its angular size (radius $\sim 4'$) is smaller than that of G321.9$-$1.1, indicating that it is a distinct object and it is opposite to the northern direction from the diffuse X-ray emission seen in the eROSITA data.

A distance of $3.29 \pm 0.75$ kpc has been reported for this SNR by \citet{Wang2020}, where the uncertainty is estimated from the half-maximum half-width of the extinction gradient profile. However, this estimate is assigned the lowest reliability class, indicating a low-confidence determination, and should therefore be treated with caution. However, if this distance is correct, it would correspond to a physical diameter of \(28.7 \pm 6.5\,\mathrm{pc}\) for the remnant.

A pulsar, PSR~J1524$-$5819, is located within the projected boundaries of the remnant. Its characteristic spin-down age is $\sim1.2\times10^5$ yrs. If this spin-down age corresponds to the pulsar's historical age, and if it would be associated with G319.8-2.0, it would define the age of the remnant. The mean radio lifetime of SNRs is expected to be about $6 \times 10^4$ years \citep{1994ApJ...437..781F} so that an age of $10^5$ years would still be consistent with what is expected for SNRs if all uncertainties involved in the lifetime estimates are taken into account. On the other side, the spin-down age of pulsars reflects the true age only if a pulsar was born with an initial period much smaller than its current period \citep{1969ApJ...157.1395O}. However, it can be used to assess the pulsar's age upper limit: \(t_{\text{max}} = 2 t_{\text{spin-down}} \approx 2.4 \times 10^5\) years \citep{2017ApJ...846..170T}. In any case, an SNR-Pulsar relation which is comparable in age is the SNR CTB 80 of which its pulsar has a spin-down age of $\sim 1.1 \times 10^5$ yrs\footnote{\url{https://www.atnf.csiro.au/research/pulsar/psrcat/}} \citep{2005AJ....129.1993M}. The pulsar's dispersion-measure-based distance listed in the ATNF pulsar catalogue is $\sim 7.5$\, kpc using the YMW16 \citep{ymw17} model. 
Since no source-specific uncertainty is provided, we adopt a characteristic fractional uncertainty of \(45\%\) for the DM-based distance estimate \citep{ymw17, 2017MNRAS.468.3289Y}, which gives \(d = 7.5 \pm 3.4\,\mathrm{kpc}\). Unfortunately, there is no other independent distance measurement for this pulsar.
Assuming the pulsar is associated with the remnant, this distance implies a physical diameter of \(67 \pm 30\) pc. This value lies within the typical range of SNR sizes; for example, studies of supernova remnants in the Magellanic Clouds find a distribution with a cut-off at \(\sim 60\) pc, while the largest remnants reach sizes slightly above \(\sim 100\) pc (e.g. J0453.9$-$7000, J0529.1$-$6833, J0536.2$-$6912; \citealt{2010MNRAS.407.1301B}).

Assuming that the pulsar is physically associated with the remnant, and that its characteristic age and DM-based distance are representative of the true system parameters, we test whether a standard SNR evolutionary model can reproduce the observed size. 

We modelled the remnant evolution using the \textsc{SNRPy} code of \citet{Leahy2017}, which provides solutions for SNR evolution.
The model requires parameters of both explosion and ambient medium parameters. The explosion is described by the kinetic energy $E_0 \approx 10^{51}\,\mathrm{erg}$, the ejected mass $M_{\rm ej}$ (in units of $M_\odot$), and the power-law index $n$ of the ejecta density profile, such that $\rho_{\rm ej} \propto r^{-n}$. The surrounding medium is parametrized by a density power-law index $s$, where $s=0$ corresponds to a uniform interstellar medium (ISM) and $s=2$ represents a stellar wind profile. In the $s=0$ case, the relevant quantity is the ISM number density $n_0$, whereas for $s=2$ the key parameters are the progenitor’s mass-loss rate $\dot{M}$ and wind velocity $v_{\rm w}$.  

We find that a set of parameters \(M_{\rm ej}=5\,M_\odot\), \(n=7\), \(s=0\), and an ambient density of \(0.6\,\mathrm{cm^{-3}}\) reproduces a remnant diameter of \(\sim 67\) pc at an age of \(1.2\times10^5\) yr. These values are within the range expected for core-collapse SNRs evolving in a uniform interstellar medium. This shows that the observed size and pulsar age could be naturally explained within a standard evolutionary scenario, consistent with the plausibility of the pulsar–remnant association. However, these results are illustrative and are not uniquely constrained because of the large distance uncertainty.

No source is detected at the position of PSR~J1524$-$5819 in the eRASS:4 data. To assess whether this non-detection is expected in eRASS:4, we estimate the X-ray flux based on the pulsar spin-down power of $\dot{E} \approx 5.6 \times 10^{33}\ \mathrm{erg\,s^{-1}}$. 
\citet{1997A&A...326..682B} reported an empirical relation between X-ray luminosity and spin-down power of the form  \( L_X (0.1\text{--}2.4~\mathrm{keV}) \sim 10^{-3} \dot{E} \). 
To obtain a conservative estimate, we adopt a possible lower efficiency of \( L_X \sim 10^{-4} \dot{E} \) when estimating the expected X-ray luminosity:
\begin{equation}
    L_X (0.1\text{--}2.4~\mathrm{keV}) \sim (10^{-4} - 10^{-3}) \dot{E},
\end{equation}
from that, we find
\begin{equation}
    L_X(0.1\text{--}2.4~\mathrm{keV}) \sim (5.6 \times10^{29} - 5.6 \times10^{30})\ \mathrm{erg\,s^{-1}}.
\end{equation}
Assuming a distance of $d = 7.5 \pm 3.4$ kpc, this corresponds to an expected flux of
\[
F_X (0.1\text{--}2.4 \text{ keV}) \approx (3.9\times10^{-17} - 2.8\times10^{-15})\ \mathrm{erg\,cm^{-2}\,s^{-1}}.
\]
To compare the expected flux with the eROSITA sensitivity, we convert it to the corresponding $0.2$--$2.3$~keV energy band used by the upper limit server\footnote{\url{https://erosita.mpe.mpg.de/dr1/AllSkySurveyData_dr1/UpperLimitServer_dr1/}} \citep{2024A&A...682A..35T} and adopt the same absorbed power-law model with photon index $\Gamma \approx 2$ and hydrogen column density \(N_{\mathrm{H}} = 3\times10^{20}\ \mathrm{cm^{-2}}\). This yields an expected absorbed flux in the range
\[
F_X(0.2\text{--}2.3~\mathrm{keV}) \approx (2\times10^{-17} - 2\times10^{-15})\ \mathrm{erg\,cm^{-2}\,s^{-1}}.
\]
To compare this with the eROSITA sensitivity, we use the eROSITA upper limit server. Since the server provides limits for eRASS1, we rescale the value to eRASS:4 by dividing by two, yielding an approximate sensitivity of
\(
 2.5\times10^{-14}\ \mathrm{erg\,cm^{-2}\,s^{-1}}.
\)
Thus, the expected flux lies below the detection threshold, and the pulsar is not expected to be detected in eRASS:4, consistent with the observed non-detection.

\subsection{MSH 15-52 (G320.4-1.2)}

MSH~15-52 (G320.4-1.2) is a young composite SNR hosting the energetic pulsar PSR~B1509–58 and its pulsar wind nebula (PWN). The PWN dominates the X-ray and $\gamma$-ray emission and gives rise to the ``cosmic hand'' morphology seen in Chandra images (see, e.g., Fig.~1 in \citealt{2025ApJ...989..221Z}). 
The remnant has an angular extent of about \(32'\) and shows a north–south asymmetry, with bright radio and X-ray emission in the northern region coincident with the optical nebula RCW~89. 
The pulsar has a characteristic age of \(\sim 1.7\)~kyr making MSH~15-52 one of the youngest known Galactic SNRs. 

Previous radio observations show synchrotron emission with a spectral index of approximately $-0.45$ \citep{1993MNRAS.264..853M}. Early X-ray observations with Einstein and ROSAT detected extended diffuse emission surrounding the pulsar, PWN, and RCW~89, with an extent similar to that seen in radio \citep{1983ApJ...267..698S,1996A&A...306..581T}. Later observations with BeppoSAX and Chandra focused mainly on the pulsar and PWN, revealing a hard, non-thermal component extending to high energies and the detailed structure of the PWN \citep{2001A&A...380..695M,2002ApJ...569..878G,2009PASJ...61..129Y,2020ApJ...895L..32B}.
At TeV energies, MSH~15-52 is associated with the $\gamma$-ray source HESS~J1514--591, whose elongated morphology broadly follows the X-ray PWN and jet structure \citep{2005A&A...435L..17A,2018A&A...612A...1H}.

These previous studies provided detailed description of the PWN, but the large-scale diffuse emission associated with the remnant remained less well constrained. The eROSITA all-sky survey data provide a wide-field view of this emission and allow us to compare the large-scale X-ray morphology with the radio structure and with the earlier ROSAT detection.

\begin{figure*}[h!]
    \centering
    \includegraphics[width=\textwidth]{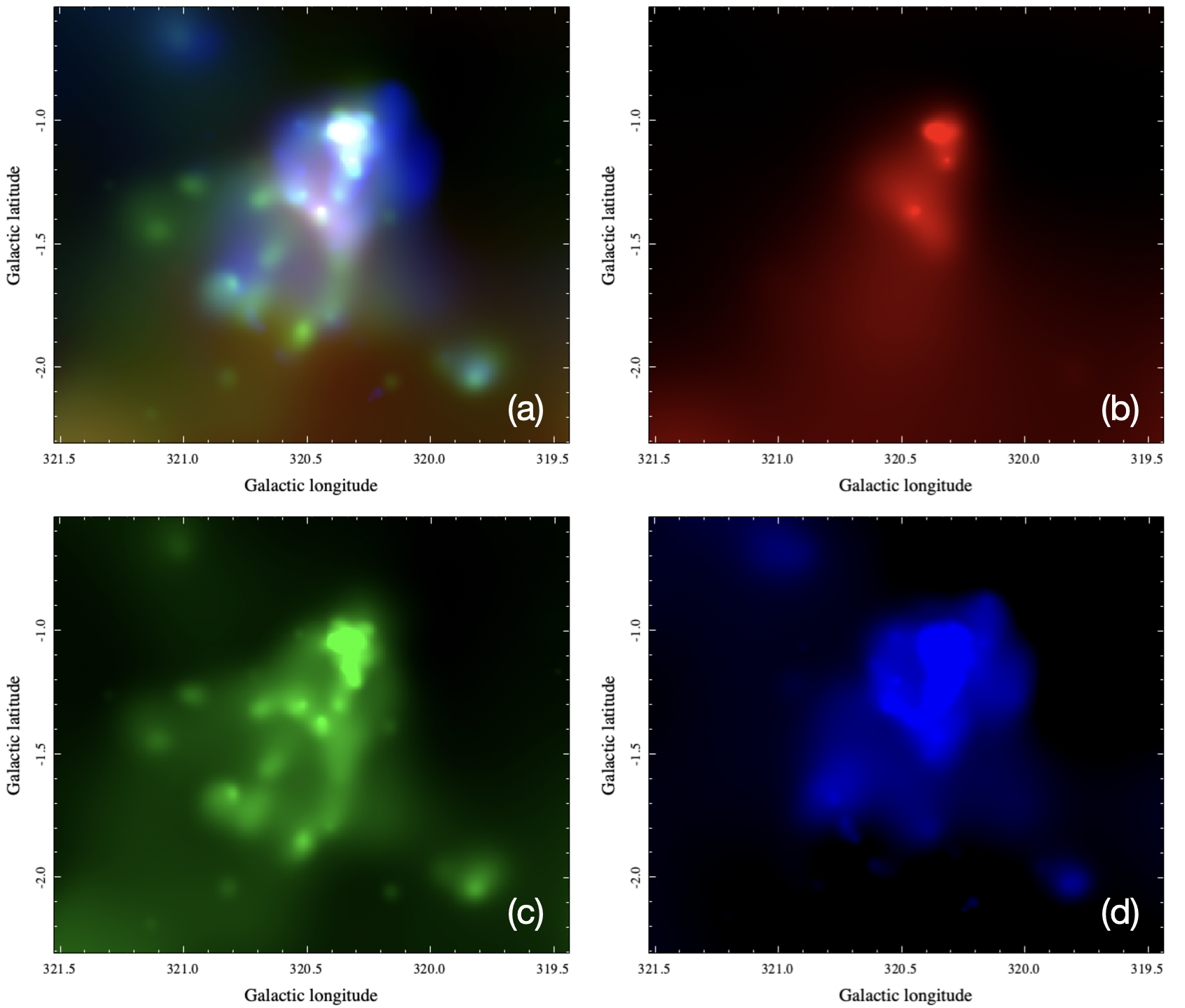}
    \caption{eROSITA eRASS:4 exposure-corrected X-ray images of MSH 15-52 and G320.6-1.6 in three energy ranges. Point sources were manually removed to enhance the visibility of the diffuse emission. (a) Combined RGB image. (b)-(d) Individual RGB eROSITA X-ray images, with the same energy bands as in Fig.~\ref{fig:all_smooth_corr}.
    Adaptive kernel smoothing was applied to each energy band, with a minimum signal-to-noise ratio of 3. 
    The colour scale was set to ASINH with the limits as follows: R: $(0.1-1) \times 10^{-3}$, G: $(0.09-3) \times 10^{-3}$, B: $(0.8-4) \times 10^{-3}$. The limits were optimized to highlight different morphological features, particularly soft emission at the intersection of the remnants and the 
    harder large-scale diffuse emission around the pulsar in MSH 15-52. 
    }
    \label{fig:MSH_colours}
\end{figure*}

\begin{figure}
    \centering
    \includegraphics[width=\linewidth]{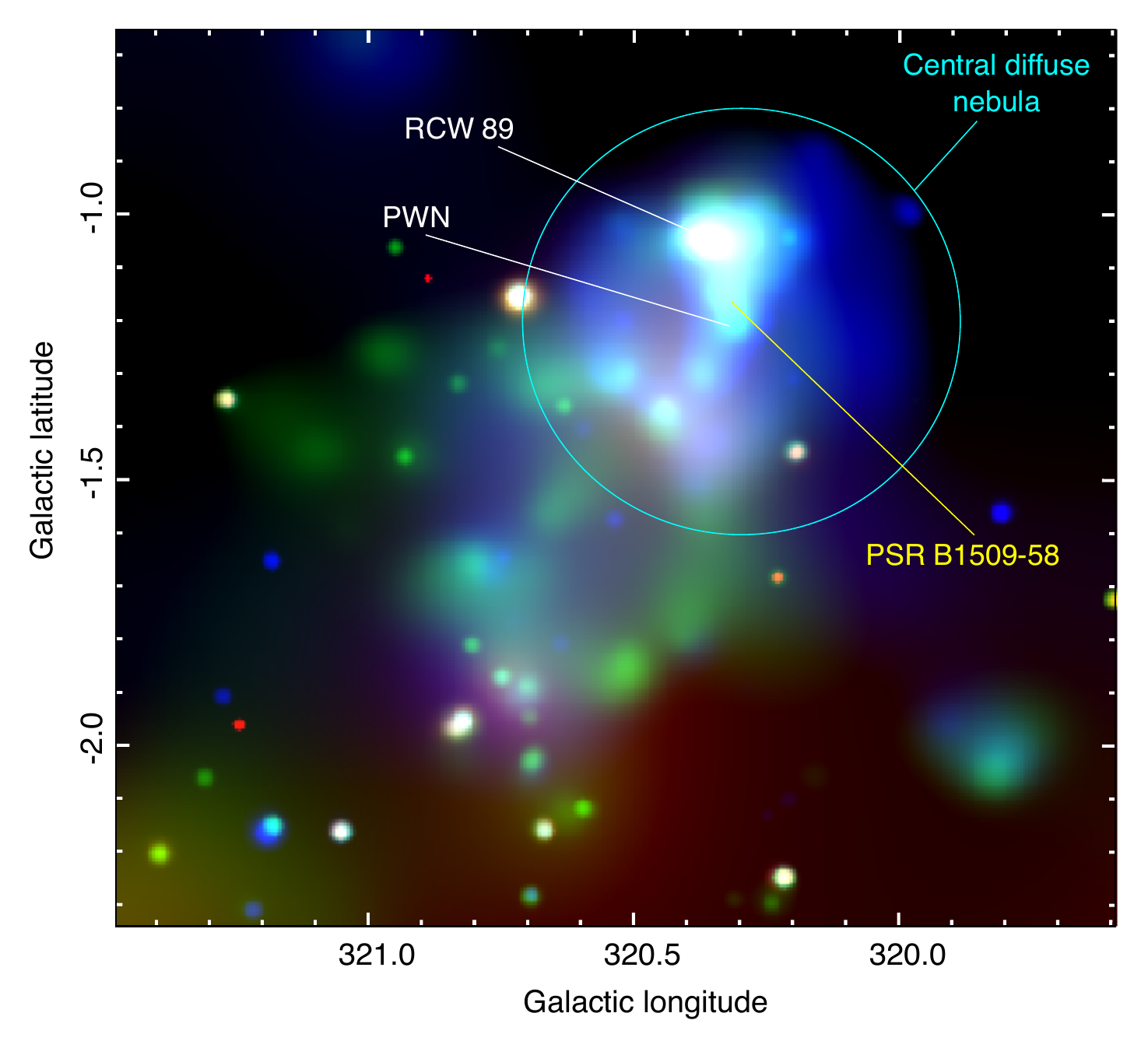}
    \caption{Adaptively smoothed eRASS:4 image showing the main structural components of MSH 15-52.}
    \label{fig:MSH_structure}
\end{figure}

\subsubsection{Spatial analysis}

The X-ray morphology of MSH~15-52 is asymmetric and consists of several distinct components. Two bright compact nebulae are visible (Fig.~\ref{fig:MSH_structure}): one centred on the pulsar, corresponding to the PWN, and another to the north, associated with RCW~89. These structures are embedded in fainter and more symmetric diffuse emission, which, following \cite{1996A&A...306..581T}, we refer to as the central diffuse nebula. This component is approximately centred on the pulsar and has a radius of $\sim 15'$.

Figure~\ref{fig:MSH_colours} shows the eRASS:4 exposure-corrected X-ray images of this field, colour-coded by photon energy (red: 0.2–0.7~keV; green: 0.7–1.1~keV; blue: 1.1–8.0~keV) and compared to radio data. eROSITA reveals diffuse X-ray emission extending over the full extent of the remnant, with emission detected up to energies of $\sim 4$~keV. While ROSAT had previously reported this extended component, its sensitivity was limited to softer energies ($E \lesssim 2.4$~keV). The higher sensitivity of eROSITA now provide a more complete view, confirming the presence of extended  emission beyond the PWN. 

\subsubsection{Distance estimates}
\label{sec:dist}

The distance to MSH~15-52 is uncertain, with published estimates ranging from $\sim 2.6$ to $5.2$~kpc (Table~\ref{tab:distances}). Since the distance sets the physical size of the remnant, we briefly summarize the available constraints and test their consistency with the observed size and age of the system using the \textsc{SNRPy} evolutionary model calculator \citep{Leahy2017}.

\begin{table*}
\centering
\caption{Distance estimates to MSH~15-52 / PSR~B1509--58 from the literature.}
\label{tab:distances}
\begin{tabularx}{\textwidth}{l c X c}
\toprule
Method & Distance, kpc & Notes & Reference \\
\midrule
H\,I absorption & $5.2 \pm 1.4$ & ATCA; most commonly adopted estimate & \citet{Gaensler1999} \\[1ex]%
Dispersion measure & $\sim 4.5 \pm2.0$ & YMW16 model; ATNF catalogue & \citet{ymw17} \\[1ex]%
Bias-corrected & $ 4.4^{+1.3}_{-0.8}$ & Combines H\,I limits, Galactic distribution, and radio luminosity priors & \citet{2012ApJ...755...39V} \\
Optical extinction & $3.0 \pm 0.5$ & Reliability class ``C'' (lowest confidence) & \citet{Wang2020} \\
$\Sigma$--$D$ relation & $\sim 2.6$ & Empirical surface-brightness--diameter relation & \citet{1993MNRAS.264..853M} \\
X-ray absorption ($N_{\rm H}$) & $> 3.9$& Based on comparison with Gaia extinction & {\bf This work }\\
\bottomrule
\end{tabularx}
\end{table*}

The most commonly adopted distance of MSH~15-52 (e.g. \citealt{2025JApA...46...14G,2025ApJ...989..221Z,2020ApJ...895L..32B}) comes from H\,I absorption studies of the surrounding radio shell, which results in $d=5.2 \pm 1.4$~kpc based on ATCA spectral line data \citep{Gaensler1999}. 
This value is broadly consistent with the dispersion measure distance of PSR~B1509--58 and with the bias-corrected estimate of \citet{2012ApJ...755...39V}. Smaller values have been inferred from optical extinction and from the empirical $\Sigma$--$D$ relation, but these estimates are less secure. In particular, the optical-extinction estimate was assigned reliability class ``C'' by \citet{Wang2020}, indicating a low-confidence determination.

Additional observational properties provide important cross-checks on the distance estimates. The characteristic spin-down age of PSR B1509--58 is $\tau_{\rm c} \approx 1700$~yr, derived from its period of 152\,ms and spin-down rate of $\dot{P} \sim 1.5 \times 10^{-12}$\,s\,s$^{-1}$ \citep{psj+19}. 
This value coincides with the epoch of the historical supernova observed in China in AD\,185, suggesting a potential association between the recorded event and MSH~15-52 \citep{1994MNRAS.268L...5S}. 

Recently, \citet{2020ApJ...895L..32B} measured blast-wave velocities of $(4000 \pm 500)\,d_{5.2}$~km\,s$^{-1}$, where $d_{5.2}$ is the distance in units of 5.2~kpc. Interpreted within a simple Sedov–Taylor framework, such velocities would imply an unrealistically young age of only a few hundred years. 

To cross-check the distance estimates with the age and size of MSH~15-52, we employed the \textsc{SNRPy} code (see Section~\ref{sec:snrpy_G321.9} for a detailed description of the model's parameters).  
The choice of ambient profile influences the modelled evolution. For a uniform medium ($s=0$), it is difficult to reproduce the large present-day radius of MSH~15-52, $R \approx 24~\mathrm{pc}$ (assuming $d=5.2~\mathrm{kpc}$), without resorting to unrealistic parameters. By contrast, a wind-shaped environment ($s=2$) provides a more natural explanation, and is also physically motivated given that MSH~15-52 is a young core-collapse remnant likely still in the ejecta-dominated stage. 

Adopting a wind profile with canonical values $E_0 = 10^{51}\,\mathrm{erg}$, $M_{\rm ej} = 15\,M_\odot$, $n = 9$, $\dot{M} = 10^{-7}\,M_\odot\,\mathrm{yr}^{-1}$, and $v_{\rm w} = 3000\,\mathrm{km\,s^{-1}}$ allows the observed radius at $d=5.2$ kpc to be reproduced. However, the corresponding forward-shock velocity in this case is $v_{\rm s} \sim 1.2\times 10^4\,\mathrm{km\,s^{-1}}$, significantly larger than the value of $\sim 4000\,\mathrm{km\,s^{-1}}$ inferred by \citet{2020ApJ...895L..32B}. Reconciling both the observed radius and shock velocity simultaneously thus remains challenging within the framework of simple analytic models. 

We further performed X-ray spectral analysis (see the following subsection for details), from which we derived the absorbing column density, $N_{\rm H}$. This value provides a lower limit on the distance of $\sim$3.9 kpc, leaving only the three largest distance estimates in Table~\ref{tab:distances} as viable. 

\subsubsection{Spectral analysis}
\label{sec:MSH_spectrum}

\begin{figure}
    \centering
    \includegraphics[width=\columnwidth]{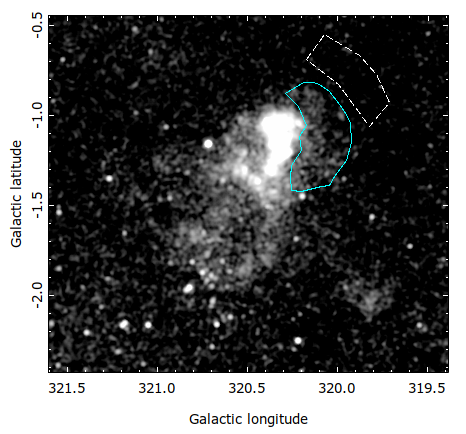}
    \caption{
    Exposure-corrected eRASS:4 image of MSH~15–52 in the 0.5–3.0 keV energy band, smoothed with a Gaussian kernel ($\sigma = 1.5$) to enhance diffuse emission. A single energy band is shown, chosen to maximize the visibility of the central diffuse nebula and to allow it to be separated from other components. The image is displayed using an ASINH colour stretch with cut values of $(0.7$–$7)\times10^{-3}$ counts s$^{-1}$. The solid cyan contour marks the source region used for the spectral analysis of the central diffuse nebula in MSH~15–52, while the dashed white contour indicates the background region. Although the diffuse emission appears largely symmetric, only the western side was used to minimize contamination from the PWN and from G320.6–1.6, which are more prominent toward the east. The spectral results are discussed in Sect.~\ref{sec:MSH_spectrum}.
    }
    \label{fig:blue-region}
\end{figure}

We analysed the emission from the central diffuse nebula, aiming to minimize contamination from the pulsar/PWN, the bright northern nebula RCW~89, and the neighbouring remnant G320.6--1.6. 
To achieve this, we restricted the analysis to the western part of the central diffuse nebula, as the eastern side is expected to suffer from contamination by these nearby structures. The extraction regions for both the source and the background are shown in Fig.~\ref{fig:blue-region} and was defined using \textsc{SAOImage DS9} \citep{2003ASPC..295..489J}. 

Previous analysis of ROSAT data modelled the central diffuse nebula's spectrum with a thermal plasma model \citep{1996A&A...306..581T}. Motivated by these results, we began by testing a set of single-component absorbed spectral models, using the \texttt{TBabs} absorption model \citep{2000ApJ...542..914W}. These included \texttt{vapec}, \texttt{pshock}, \texttt{nei}, and a power-law. The \texttt{vapec} model provided an acceptable statistical description of the spectrum (with the goodness of 29\%), with a best-fit temperature of $kT = 0.78^{+0.08}_{-0.07}$~keV and sub-solar abundances. However, the derived hydrogen column density ($N_{\rm H} \sim 1.8_{-0.3}^{+0.8} \times 10^{22}$~cm$^{-2}$) exceeds or is close to (considering the lower limit) the total Galactic column in this direction ($N_{\rm H} = 1.6 \times 10^{22}$~cm$^{-2}$; \citealt{2013MNRAS.431..394W}) and is poorly constrained, leading us to disfavour this model.  

Non-equilibrium models (\texttt{pshock}, \texttt{nei}), which might be expected for a young ($\sim 2$~kyr) remnant, did not perform well: they yielded implausible parameters and no statistical improvement over the equilibrium model. A pure power-law  provided a statistically comparable fit to \texttt{vapec} (goodness $\approx 27\%$), with $\Gamma = 1.6$ and $N_{\rm H} \sim 0.5 \times 10^{22}$~cm$^{-2}$.

We also tested a composite thermal+nonthermal model. The \texttt{powerlaw+vapec} fit provided the best overall description (minimal AIC), with $kT = 0.77^{+0.08}_{-0.09}$~keV, $\Gamma = 1.5 \pm 0.2$, and $N_{\rm H} = 0.63_{-0.14}^{+0.15} \times 10^{22}$~cm$^{-2}$. 
This absorption value is consistent within \(2 \sigma\) error range with the independent estimate of $(9.5 \pm 0.3) \times 10^{21}$~cm$^{-2}$ by \citet{2002ApJ...569..878G}. 
The composite model simultaneously reproduces the thermal line emission and the harder continuum, making it our preferred description of the central diffuse nebula spectrum. The results of all fits are summarized in Table~\ref{tab:MSH-spectrum}. 

Adopting an absorbing column density of $N_{\rm H} = 7 \times 10^{21}$ cm$^{-2}$ and using the same method as described in Sect.~\ref{sec:dist}, we derive an extinction value of $A_V \simeq 3.4$. However, available Gaia extinction measurements along this line of sight extend only up to $A_V \approx 2.2$, which corresponds to a distance of approximately 3.9 kpc which we therefore consider a lower limit on the distance to MSH 15-52.

The absorbing column density obtained with the preferred composite model,
\(N_{\mathrm{H}} = 0.63^{+0.15}_{-0.14}\times10^{22}\,\mathrm{cm^{-2}}\), is lower than the value reported from the ROSAT analysis of the central diffuse nebula by \citet{1996A&A...306..581T}, who obtained \(N_{\mathrm{H}} = 1.45^{+0.06}_{-0.04}\times10^{22}\,\mathrm{cm^{-2}}\) for a thermal model. However, this comparison should be treated with caution because the ROSAT result was obtained with lower photon statistics and because the inferred column density depends on the adopted spectral model. In the same ROSAT study, a purely non-thermal model also provided an acceptable fit, but yielded a much lower column density of \(N_{\mathrm{H}} = 0.30^{+0.09}_{-0.10}\times10^{22}\,\mathrm{cm^{-2}}\). Thus, the ROSAT data alone do not uniquely constrain the absorbing column density of the central diffuse nebula. 

The temperature of the thermal component, \(kT = 0.77^{+0.08}_{-0.09}\,\mathrm{keV}\), is somewhat higher than the ROSAT value of \(kT = 0.53^{+0.05}_{-0.06}\,\mathrm{keV}\), but the two estimates are still consistent within the \(2\sigma\) uncertainties. The non-thermal component is also consistent with the ROSAT power-law fit, for which \citet{1996A&A...306..581T} found \(\Gamma = 1.7\pm0.4\), compared to \(\Gamma = 1.5\pm0.2\) in our analysis.

\begin{table*}[h!]
\centering
\caption{The results of the spectral analysis of the central diffuse nebula inside MSH 15-52. Best fit parameters with $1 \sigma$ errors obtained with different models with Ne, Mg, Si, and Fe free to vary. 
\label{tab:MSH-spectrum}}
\begin{tabular}{cccc}
\toprule
Model & TBabs*vapec   & TBabs*powerlaw & TBabs*(powerlaw+apec) \\
\midrule
$N_\text{H}$ (10$^{22}$ cm$^{-2}$) &  $1.8_{-0.3}^{+0.8}$ &  $0.53_{-0.09}^{+0.12}$ & $0.63_{-0.14}^{+0.15}$\\[1ex]%
kT (keV)  & $0.78_{-0.07}^{+0.08}$ & - & $0.77_{-0.09}^{+0.08}$\\[1ex]%
Ne/Ne$_{\odot}$ & $ < 1$ & -  & - \\[1ex]%
Mg/Mg$_{\odot}$ & $0.3_{-0.2}^{+0.4}$ & - & - \\[1ex]%
Si/Si$_{\odot}$ & $ <0.11  $ & & \\[1ex]%
Fe/Fe$_{\odot}$ &  $0.3_{-0.2}^{+0.7}$ & - & - \\[1ex]%
Norm$_{\text{therm}}$ &  $0.029_{-0.006}^{+0.011}$  &  -  &  $0.0008_{-0.0005}^{+0.0011}$ \\[1ex]%
$\Gamma$ & -  & $1.6_{-0.2}^{+0.3}$ & $1.5_{-0.2}^{+0.2}$  \\[1ex]%
Norm$_{\text{PL}}$ & -  & $0.0025_{-0.0003}^{+0.0004}$ & $0.0025_{-0.0004}^{+0.0003}$ \\[1ex]%
\midrule
Statistic / d.o.f. & 1802/1603 & 1795/1607 & 1788/1605 \\ 
goodness & 29\% & 27\% & 23\% \\
AIC & 1856 & 1841 & 1838 \\
\(\Delta\)AIC & 18 & 3 & 0 \\
\bottomrule
\end{tabular}
\end{table*}

\subsection{G320.6-1.6}

G320.6--1.6 is a known SNR located close to MSH 15–52, partially overlapping with it in the west.
The remnant was identified in radio with the Parkes 64-m telescope as a large, low-surface-brightness structure with several arcs which are predominantly polarized tangentially, distinct from the radial polarization pattern observed in MSH 15-52 \citep{1993MNRAS.264..853M}. 
Despite its proximity to the well-studied MSH 15-52, G320.6-1.6 remains much less explored, with prior studies focusing mostly on its radio and optical properties. 
In contrast to MSH~15--52, no central compact object or pulsar has been identified in association with G320.6--1.6. 

Using the Parkes 64-m telescope at 4.8 and 8.4~GHz, \cite{1993MNRAS.264..853M} studied the faint, polarized radio shell corresponding to G320.6--1.6. 
At lower frequency and much higher resolution, the 843~MHz Molonglo Observatory Synthesis Telescope (MOST) images of \citet{1996A&AS..118..329W} revealed an incomplete shell composed of multiple arcs and filaments.

G320.6-1.6 has an angular size of approximately $\sim 1.0^{\circ} \times 0.7^{\circ}$ (see Fig~\ref{fig:X+radio}).
The radio emission is non-thermal, with a spectral index of $\alpha = -0.47$ \citep{1974A&AS...18..267G}, typical for synchrotron emission from SNRs. 

The source was not detected in the Einstein IPC observations of the region, but it was later identified as a faint X-ray feature in the ROSAT PSPC data, where it was referred to as the ``South-East Nebula'' (\citealt{1996A&A...306..581T}). Fixing the column density to $N_{\mathrm{H}} = 1.27\times10^{22}\ \mathrm{cm^{-2}}$, \citet{1996A&A...306..581T} obtained a best-fit plasma temperature of $kT \approx 0.17$~keV, indicating very soft emission. 

A $\gamma$-ray source listed in the incremental \emph{Fermi}-LAT fourth source catalogue (4FGL-DR3; \citealt{2022ApJS..260...53A}), 4FGL~J1518.9$-$5903c, is reported as likely associated with SNR~G320.6$-$1.6. The catalogue gives a Bayesian association probability of 0.80, and the 95\% localization ellipse ($0.178^{\circ}\times0.103^{\circ}$) lies completely within the boundaries of the remnant. The source is detected with an average significance of 4.33$\sigma$ and is best described by a power-law spectrum with photon index $\Gamma = 2.84 \pm 0.16$. 
However, given the ``c'' designation, which indicates possible systematic effects related to the modelling of Galactic diffuse emission, this association should be treated cautiously.

\subsubsection{Spatial analysis}

\begin{figure}
    \includegraphics[width=\linewidth]{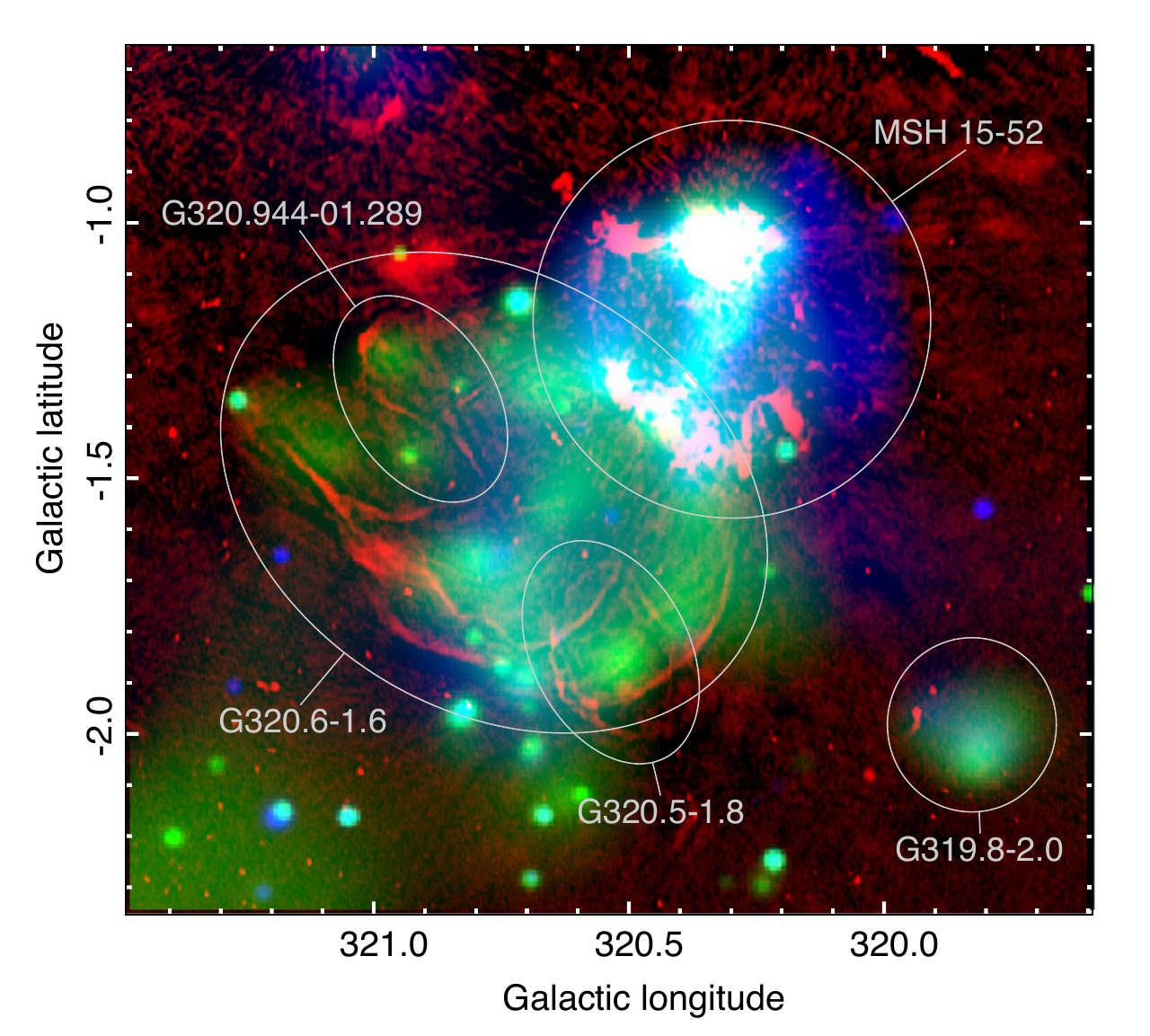}
    \caption{Combined radio and X-ray image of MSH~15-52 and G320.6-1.6. Red shows the RACS-low 888~MHz radio emission. Green and blue represent eROSITA eRASS:4 exposure-corrected X-ray images, colour-coded by energy: green corresponds to 0.7–1.1~keV and blue to 1.1–8.0~keV. Adaptive kernel smoothing was applied to each X-ray band, with a minimum signal-to-noise ratio of 3. }
    \label{fig:X+radio}
\end{figure}

Fig.~\ref{fig:X+radio} presents a composite X-ray+radio image of MSH 15-52 and G320.6-1.6. In this image, the radio emission is represented in red, using data from the RACS-low (Rapid ASKAP Continuum Survey, 888~MHz) survey \citep{2020PASA...37...48M, 2021PASA...38...58H}. The green and blue channels are the same as in Fig.~\ref{fig:all_smooth_corr}, but with adjusted colour scale limits: $(0.3-5) \times 10^{-6}$ for green and $(0.4-2) \times 10^{-6}$ for blue.

As shown in Fig.~\ref{fig:MSH_colours}, the emission is detected predominantly in the medium-energy band (0.7--1.1~keV).
The overall morphology is elongated, with hints of a partial shell-like outline, although the shell is not continuous and does not form a closed rim. 

The radio structure in the north-east shows an unusual morphology, somewhat reminiscent of a radio lobe or jet-like feature. It is listed in the catalogue of SMGPS SNR candidates (Table~A4; \citealt{2025A&A...693A.247A}) as G320.944$-$01.289, with a radius of $7.4'$ and reliability class of III, indicating a low-confidence candidate.
Diffuse X-ray emission is also detected near the centre of this structure. To investigate a possible extragalactic origin, we cross-matched the region with available AGN and quasar catalogues. Due to the proximity to the Galactic plane, the coverage of such catalogues is incomplete; however, no counterparts are found in the accessible databases.
\citet{2025ApJ...988...75B} suggest that this feature may be associated with the SNR G320.6$-$1.6.
 At present, the nature of this feature remains unclear.

\subsubsection{X-ray spectral analysis}
\label{sec:spectrum_G320.6_GRXE}

\begin{figure}[b!]
    \centering
    \includegraphics[width=\linewidth]{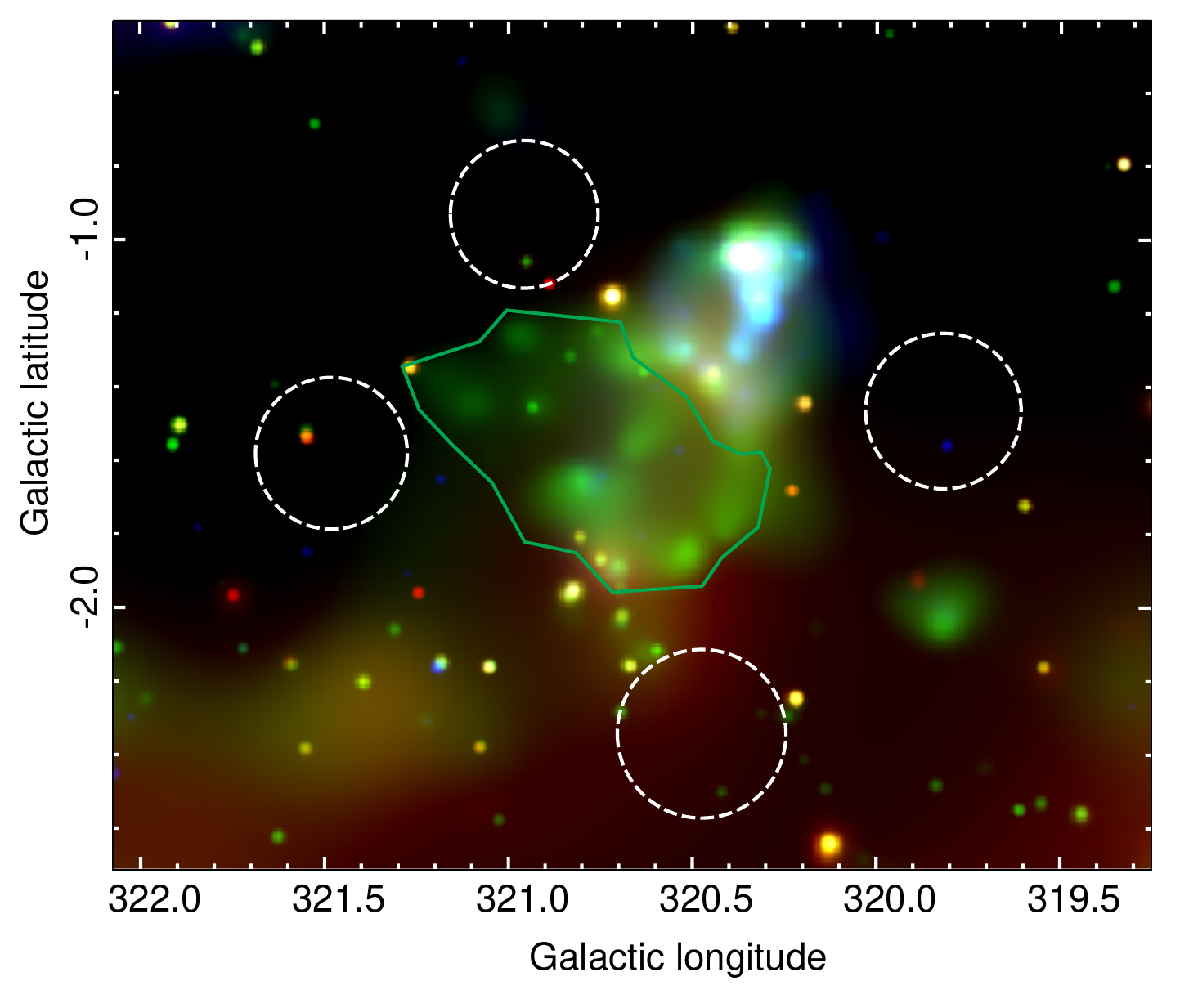}
    \caption{Regions used for the spectral analysis of G320.6--1.6. The solid green line defines the region used for the extraction of source counts. The white dashed circles indicate the background extraction regions.}
    \label{fig:G3206_regions}
\end{figure}

We performed the X-ray spectral analysis following the same procedure as described above. 
The source and background extraction regions are shown in Fig.~\ref{fig:G3206_regions}.
For the initial fits we combined all four background regions to obtain a representative estimate of the local sky background.

We tested several single-component thermal plasma models, including \texttt{vpshock}, \texttt{vapec}, and \texttt{vnei}, each multiplied by a \texttt{TBabs} absorption component. 
All models provide statistically comparable fits, as summarized in Table~\ref{tab:G320.6}, which lists the best-fit parameters for each case. 
Across all models, we derive a consistent hydrogen column density of $N_{\rm H} \simeq 1.3 \times 10^{22}$~cm$^{-2}$.
This is in agreement with the ROSAT measurement of 
\(N_{\rm H} = 1.27^{+0.13}_{-0.31} \times 10^{22}~\mathrm{cm^{-2}}\) (95\% confidence; \citealt{1996A&A...306..581T}).  
This value is comparable to the total Galactic absorption expected in this direction ($N_{\rm H} \approx 1.1 \times 10^{22}$~cm$^{-2}$; H\,I surveys\footnote{\url{https://www.astro.uni-bonn.de/hisurvey/profile/index.php}}, \citealt{2005A&A...440..775K, 1997agnh.book.....H, 2005A&A...440..767B, 2000A&AS..142...35A}).

\begin{table*}[]
\centering
\caption{Best-fit spectral parameters (with $1\sigma$ uncertainties) for SNR~G320.6$-$1.6, derived from the region defined in Fig.~\ref{fig:G3206_regions}. Results are shown for the different thermal plasma models, with the O, Ne, Mg, Si, and Fe abundances allowed to vary. 
\label{tab:G320.6}}
\begin{tabular}{cccc}
\toprule
Model & TBabs*vapec & TBabs*vnei  &  TBabs*vpshock \\
\midrule
$N_{\text{H}}$ (10$^{22}$ cm$^{-2}$) & $1.36_{-0.09}^{+0.13}$  &  $1.36_{-0.06}^{+0.09}$  & $1.35_{-0.05}^{+0.07}$\\[1ex]%
kT (keV) & $0.17_{-0.02}^{+0.02}$ &  $0.17_{-0.01}^{+0.01}$ &  $0.17_{-0.01}^{+0.01}$ \\[1ex]%
O/O$_{\odot}$ & $< 0.2$ &  $< 0.2$ & $< 0.12$ \\[1ex]%
Ne/Ne$_{\odot}$ & $0.15_{-0.03}^{+0.08}$ & $0.14_{-0.02}^{+0.07}$ &  $0.14_{-0.03}^{+0.07}$ \\[1ex]%
Mg/Mg$_{\odot}$ & $0.23_{-0.07}^{+0.11}$ & $0.23_{-0.06}^{+0.08}$ & $0.23_{-0.06}^{+0.07}$  \\[1ex]%
Si/Si$_{\odot}$  & $5_{-2}^{+6}$ & $5_{-3}^{+2}$ & $5_{-1}^{+1}$ \\[1ex]%
Fe/Fe$_{\odot}$ & $0.9_{-0.4}^{+0.9}$ & $0.9_{-0.2}^{+0.8}$ &  $0.8_{-0.2}^{+0.7}$  \\[1ex]%
Tau  (10$^{10}$ cm$^{-3}$ s)  & -  & N/A &  N/A \\[1ex]%
Normalization & $3.5_{-1.4}^{+2.2}$ & $3.7_{-0.8}^{+0.8}$ &  $3.6_{-0.7}^{+1.4}$ \\[1ex]%
\midrule
Statistic / d.o.f.  &  1671/1602 & 1670/1601  &   1670/1601 \\ %
goodness & 65\% & 68\% & 66\% \\
AIC & 1727 & 1728 & 1728 \\
\(\Delta\)AIC & 0 & 1 & 1 \\
\bottomrule \\
\end{tabular}
\end{table*}

Across all models we obtain a temperature of \(kT = (0.17 \pm 0.02)\)~keV, in agreement with the ROSAT value (\(kT = 0.17 \pm 0.02\)~keV when fixing \(N_{\rm H} = 1.27 \times 10^{22}~\mathrm{cm^{-2}}\)). A representative \texttt{vapec} fit is shown in Fig.~\ref{fig:G3206_spectr}. The spectrum displays a strong Si\,\textsc{xiii} line near 1.85~keV. However, this feature is also  present in the background spectrum and is not reproduced by our initial background model, which resulted in artificially enhanced Si abundances in the source fits.

Because the target lies close to the Galactic plane, the background is expected to include a significant contribution from the Galactic Ridge X-ray Emission (GRXE) -- the narrow band of apparently diffuse X-ray emission that is concentrated along the Milky Way. 
Previous work has demonstrated that the surface brightness of the GRXE correlates tightly with the near-infrared light of the Galaxy, implying that the emission traces the stellar mass distribution  \citep{2006A&A...452..169R}. 
Deep \textit{Chandra} observations further have shown that a large fraction of the GRXE is resolved into discrete sources such as cataclysmic variables and coronally active binaries \citep{2009Natur.458.1142R}.
The Si\,\textsc{xiii} line detected in the background spectra is therefore naturally explained as part of the GRXE foreground/background emission, reinforcing the need to include an explicit GRXE component in the background modelling.

We adopted the spectral shape derived by \citet{1997ApJ...491..638K} in their ASCA study of the GRXE, which covers a similar energy range to eROSITA and exhibits the Si K line present in our data. Their analysis used a two-component ionization nonequilibrium (NEI) plasma model. Following this approach, we added a tbabs*nei+tbabs*nei component to the sky background.

\begin{figure}
    \centering
    \includegraphics[width=\linewidth]{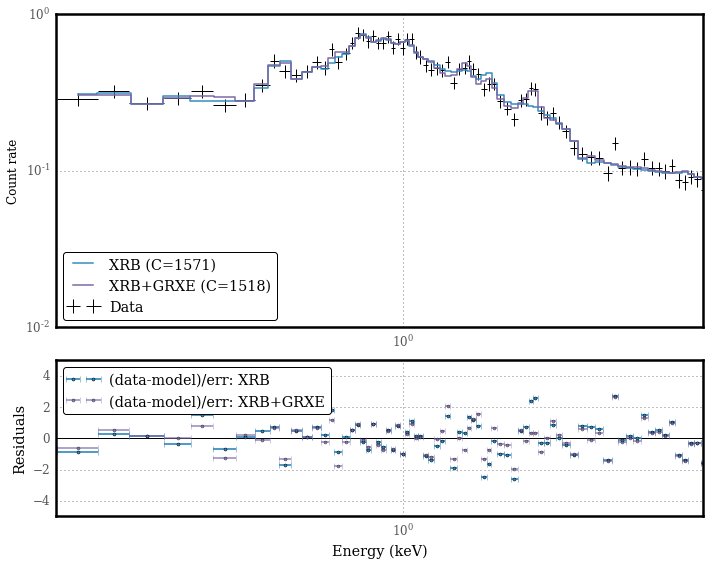}
    \caption{Comparison of the background fits obtained with (purple) and without (blue) inclusion of the GRXE component. The spectra were extracted from the four background regions shown in Fig.~\ref{fig:G3206_regions}.}
    \label{fig:GRXE_compare}
\end{figure}

Incorporating this GRXE component successfully accounts for the previously unexplained Si emission and yields an improved fit for both the background and the SNR spectra (see Fig.~\ref{fig:GRXE_compare}).

We also analysed the emission from SNR~G320.6$-$1.6 after excluding the region associated with the source G320.5-1.8 (see the details in the following Subsection~\ref{sec:WRB}), using the background model that includes the GRXE component. In this case the spectrum is well described by a single \texttt{apec} model (Table~\ref{tab:G3205}). The resulting best-fit parameters differ from those obtained when the G320.5-1.8 region was included. In particular, the inferred absorption column is lower, $N_{\rm H} \simeq 5.5 \times 10^{21}\,\mathrm{cm^{-2}}$, a value that is more consistent with the total value of Galactic absorption expected for this line of sight ($N_{\rm H} \approx 1.1 \times 10^{22}$~cm$^{-2}$). The plasma temperature also increases, reaching $kT \approx 0.7$~keV. These differences indicate that the G320.5-1.8 region may contribute distinct spectral components which, when included, bias the parameters derived for the larger remnant.

\subsection{G320.5-1.8}
\label{sec:WRB}

\begin{figure}
    \centering
    \includegraphics[width=\linewidth]{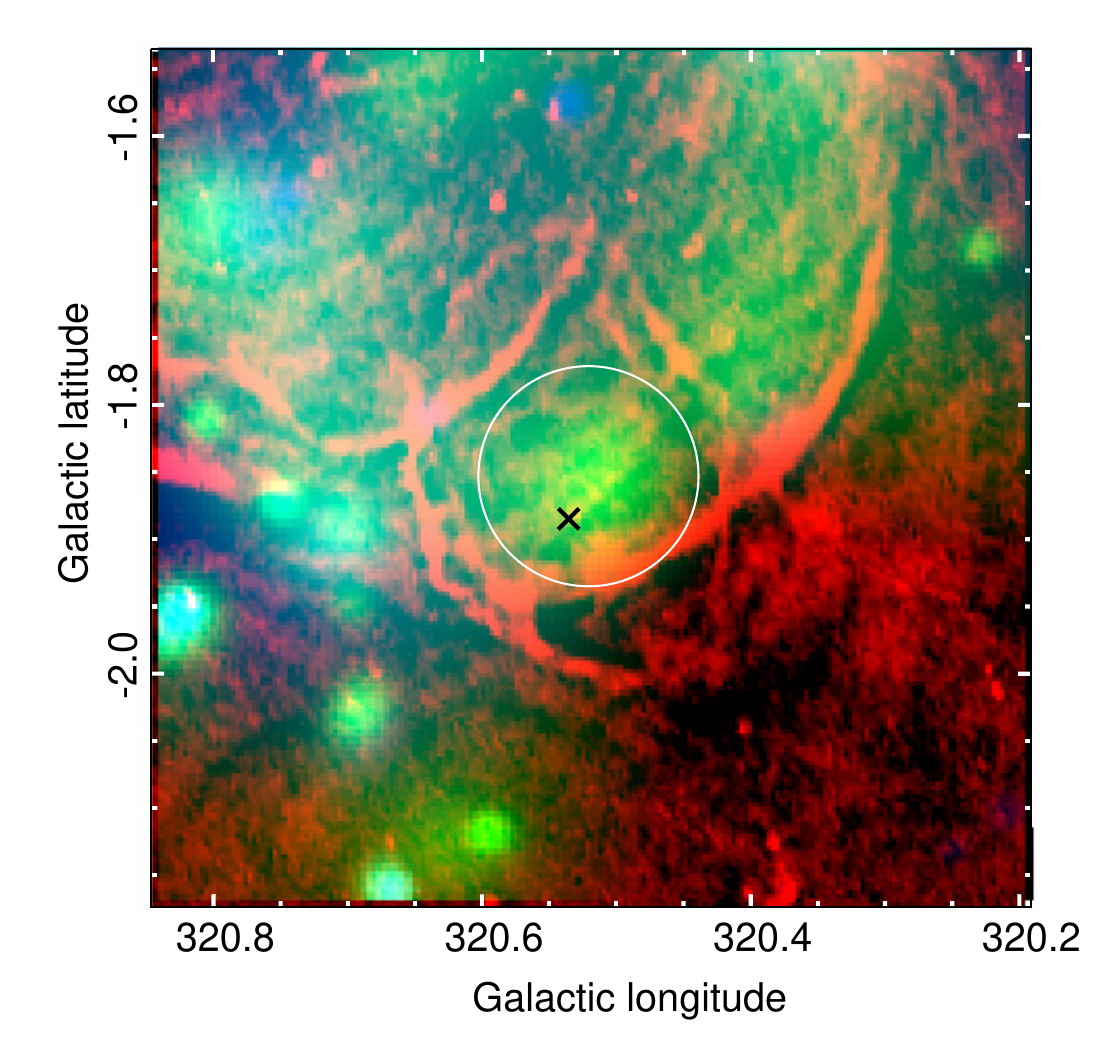}
    \caption{Same as Fig.~\ref{fig:X+radio}, but zoomed in on G320.5-1.8. The white circle indicates the region used for the spectral analysis of G320.5-1.8, and the black cross shows the position of WR star WR~68.}
    \label{fig:G320518}
\end{figure}

\begin{figure*}
    \centering
    \includegraphics[width=0.5\textwidth]{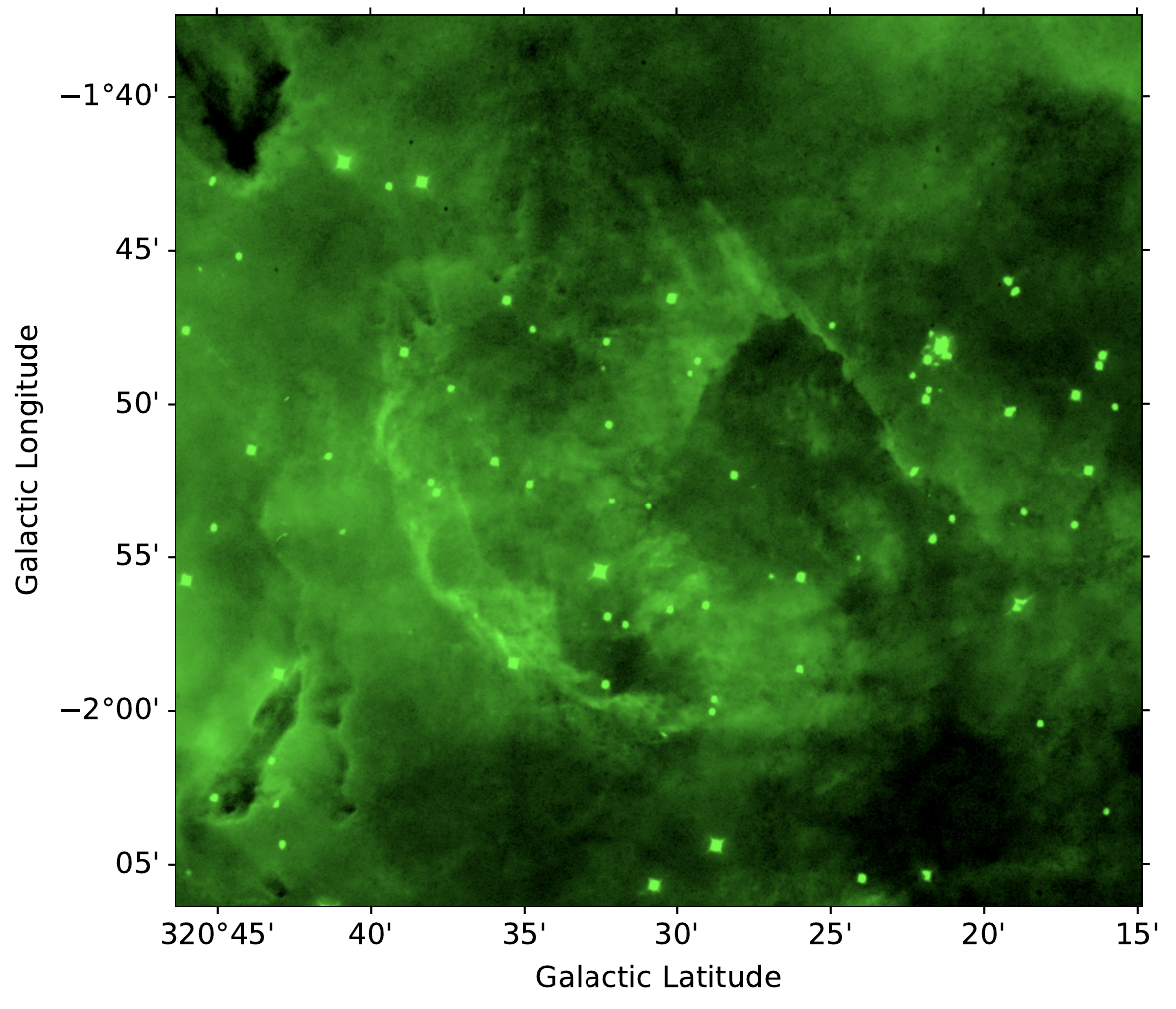}%
    \includegraphics[width=0.5\textwidth]{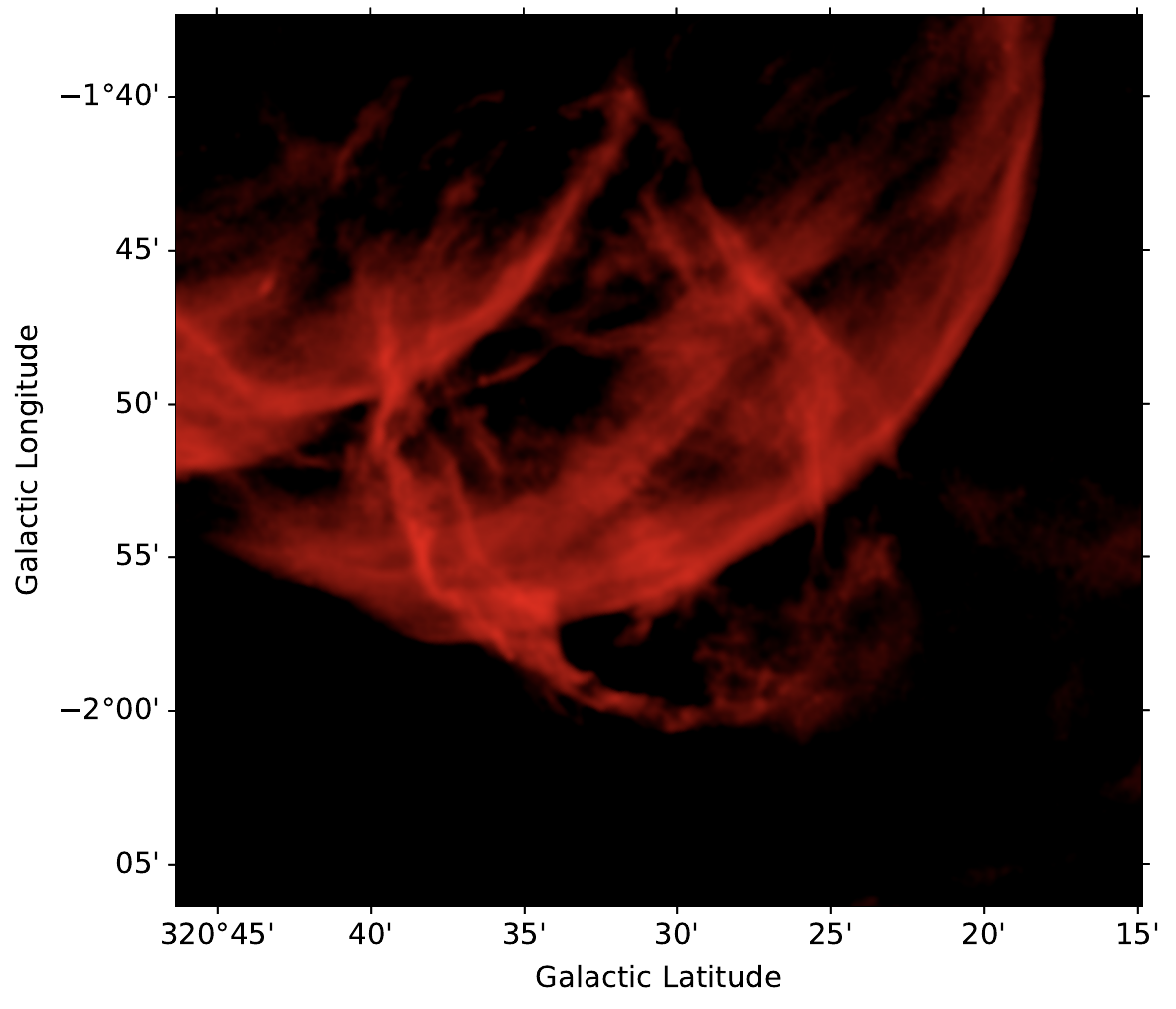}
    \caption{G320.5--1.8 as seen in the SuperCOSMOS H$\alpha$ survey (left) and EMU radio continuum emission (right). Point sources were removed from both images to the greatest extent possible, although some residual point sources remain visible in the optical image.}
   \label{fig:G3205-ha+radio}
\end{figure*}

An analysis of the radio continuum data within the boundaries of G320.6-1.6 reveals a distinct smaller shell (G320.5-1.8) in the southern part of the field (see Fig.~\ref{fig:X+radio}). This feature is nearly complete and has an elliptical morphology (see Fig.~\ref{fig:G320518}). 
Its centre is located at Galactic coordinates $l=320.5^{\circ}$, $b=-1.84^{\circ}$, with semi-major and semi-minor axes of $11.8'$ and $8.4'$, respectively. 
The shell is well defined in the radio band, and its eastern arc coincides spatially with prominent H\(\alpha\) emission (see Fig.~\ref{fig:G3205-ha+radio}). 
In addition, a region of X-ray emission is visible near the centre of the shell in the X-ray eROSITA image.

The  H$\alpha$ emission was first identified as a possible Wolf–Rayet (WR) star bubble associated with WR~68 (RA $=15^{\mathrm h}18^{\mathrm m}20.76^{\mathrm s}$, Dec $=-59^\circ38'17.3''$) by \citet{1994ApJS...93..229M}. It was later detected and included among Galactic WR nebulae by \citet{2010MNRAS.409.1429S}, and was further discussed as a WR bubble by \citet{2015A&A...578A..66T}. For an adopted distance of \(\sim 3.3\)~kpc to WR~68, \citet{1997ApJ...475..188M} estimated a physical diameter of \(\sim 13\)~pc, which is consistent with the size range of known WR bubbles. The same H\(\alpha\) emission was also noted by \citet{2011MNRAS.414.2282S}, who tentatively associated it with the SNR G320.6--1.6; however, the possible association with WR~68 was not discussed in that work.

Several arguments support the WR bubble interpretation. First, WR~68 is located within the projected shell (Fig.~\ref{fig:X+radio}). Second, the radio and H\(\alpha\) emission trace the same elliptical structure, while the rest of the radio filaments of G320.6-1.6 do not have detected optical counterparts. Third, radio continuum emission is frequently detected from WR bubbles, including examples such as those associated with WR\,7, WR\,136, WR\,40, WR\,124, WR\,16, and WR\,102. Although no complete census of Galactic WR bubbles exists, radio emission is commonly observed in prominent WR bubbles.
We compiled a working catalogue of Galactic WR bubbles by combining objects listed in previous optical and infrared surveys \citep{2010MNRAS.409.1429S, 2015A&A...578A..66T}, resulting in a sample of 40 candidate nebulae. For each object, we performed a  visual inspection of available radio continuum data, primarily from the RACS-low and RACS-mid \citep{2020PASA...37...48M, 2023PASA...40...34D}, to assess the presence of associated radio emission. Using this approach, 
we found that approximately half (19 out of 40) of the objects have a radio counterpart. We note that this classification is based on morphological inspection rather than automated detection, and that the degree of association varies across the sample, particularly in regions of complex Galactic-plane emission.

Taken together, the data favour the interpretation that G320.5--1.8 is a separate object projected within the boundaries of G320.6--1.6. 
Its association with the WR~68 bubble is therefore a plausible hypothesis. However, the current data do not provide a secure classification, and an interpretation as a separate SNR cannot be fully excluded. Optical spectroscopic observations, a radio spectral index map, and follow-up X-ray observations are required to confirm the nature of this object and to determine whether the X-ray emission is physically associated with the optical/radio shell.

\subsubsection{Spectral analysis}

We next examined the diffuse X-ray emission inside G320.5$-$1.8. Using the extraction region shown in Fig.~\ref{fig:G320518}, and adopting the same set of background regions as for G320.6$-$1.6, we extracted the spectrum and modelled it with an absorbed \texttt{apec} plasma component. Since this region has a lower number of source counts compared to G320.6$-$1.6, we did not employ the more flexible \texttt{vapec} model. The background model included the GRXE component.

The resulting spectrum is shown in Fig.~\ref{fig:G3205-spectrum}, and the best-fit parameters are listed in Table~\ref{tab:G3205}. The derived spectral properties differ from those obtained for G320.6$-$1.6. In particular, the hydrogen column density is lower,
\(N_{\rm H} = (0.8 \pm 0.2)\times 10^{22}~\mathrm{cm^{-2}},\)
and the plasma temperature is higher,
\( kT = 0.30^{+0.07}_{-0.05}~\mathrm{keV}. \)
These differences indicate that the two regions might trace distinct physical conditions or emission components, rather than belonging to a single coherent X-ray structure.

\begin{table}[]
\centering
\caption{Best-fit spectral parameters (with $1\sigma$ uncertainties) for the candidate source G320.5-1.8 and for SNR~G320.6$-$1.6 excluding the G320.5-1.8 region. Results are shown for the \texttt{apec} model, using a background that includes the GRXE component.
\label{tab:G3205}}
\begin{tabular}{ccc}
\toprule
Region &  G320.5-1.8  &  G320.6-1.6*    \\%
\midrule
$N_\text{H}$ (10$^{22}$ cm$^{-2}$) & $0.8_{-0.2}^{+0.2}$ & \(0.55^{+0.06}_{-0.07}\) \\[1ex]
kT (keV) & $0.30_{-0.05}^{+0.07}$ & \(0.71\pm0.03\) \\[1ex]
Abundance & $< 0.6$ & \(0.20^{+0.09}_{-0.05}\) \\[1ex]
Norm & $0.019_{-0.011}^{+0.023}$ & \(0.029 \pm 0.005\) \\[1ex]
\midrule
Statistic / d.o.f. & 1552/1596 & 1636/1596 \\
goodness & 40\% & 57\% \\
\bottomrule
 \end{tabular} 
 \end{table}

\section{Summary}

We have conducted a multiwavelength study of SNRs and SNR candidates in a region of the Circinus constellation, combining eROSITA X-ray survey data with radio, optical, and infrared observations. 
In the following, we summarize the findings for each object, classified based on the available multiwavelength diagnostics, which are also given in Table~\ref{tab:all}.

\begin{table*}[htb!]
\centering
\caption{Summary of the sources studied in this work.}
\label{tab:all}
\begin{tabularx}{\textwidth}{l l X X X  }
\toprule
 Name &  Radio sp. index & Optical detection & X-ray evidence & Status  \\
\midrule
MSH 15-52 & \(-0.45\) [1] & N/A & thermal and non-thermal & confirmed SNR \\
G320.6-1.6 & \(-0.47\) [2] & N/A & thermal & confirmed SNR \\
G320.2-3.6 & N/A & [S II], [OIII] [3] & thermal & promising candidate SNR\(^{**}\)\\
G321.9-1.1 & N/A & N/A & unclear & possible SNR\(^{*}\)  \\
G321.8-1.1 & N/A & N/A & unclear & possible SNR \\
G320.5-1.8 & N/A & N/A & thermal & possible SNR / possible wind-blown bubble around WR~68  \\
G319.8-2.0 & N/A & N/A & thermal & unclear \\
\bottomrule
\end{tabularx}
\caption{References: [1]: \cite{1993MNRAS.264..853M}; [2]: \cite{1974A&AS...18..267G}; [3]: \cite{2024Fesen}.\\ 
\(^{*}\)Classified as a SNR in the literature (e.g. \citealt{2025JApA...46...14G}), but lacks strong evidence. The evidence only includes the shell shape in radio and possible X-ray counterpart. \\
\(^{**}\)Was considered a ``blow-out'' of G321.3-3.9 \citep{2024Fesen}.
}
\end{table*}

\subsection{Confirmed SNRs (robust)}

\textbf{MSH~15$-$52.} 
For this known remnant, having non-thermal radio emission combined with an X-ray counterpart and associated pulsar, we confirm and study in detail the degree-scale diffuse emission, which is most prominent in the harder (B, 1.1--8.0 keV) band. The X-ray spectrum is best described by a \texttt{TBabs*(powerlaw+vapec)} model, yielding a plasma temperature of \(kT = 0.77^{+0.08}_{-0.09}\) keV, a photon index of \(\Gamma = 1.5 \pm 0.2\), and a hydrogen column density of \(N_{\rm H} = 0.63^{+0.15}_{-0.14} \times 10^{22}\ \mathrm{cm^{-2}}\).\\

\noindent \textbf{G320.6$-$1.6.} 
This object has been classified as a confirmed SNR based on its non-thermal radio emission and the detection of diffuse X-ray emission. Using eROSITA data, the X-ray emission can be reasonably described by a thermal \texttt{apec} model, yielding \(N_{\rm H} = 1.36^{+0.13}_{-0.09} \times 10^{22}\ \mathrm{cm^{-2}}\) and \(kT = (0.17 \pm 0.02)\) keV.
However, the derived spectral parameters are sensitive to the choice of extraction region. When excluding the area associated with G320.5$-$1.8 (see below), the fit yields \(N_{\rm H} = 0.55^{+0.06}_{-0.07} \times 10^{22}\ \mathrm{cm^{-2}}\) and \(kT = 0.71 \pm 0.03\) keV.
This variation may indicate the presence of spatially distinct emission components and suggests that the spectral properties should be interpreted with caution.

\subsection{Promising candidate SNR (multiwavelength support)}

\textbf{\GG.} 
We identify this object as a promising new SNR candidate. It is located right next to and partly overlaps with the known SNR  G321.3-3.9. It displays a shell-like morphology in radio continuum and corresponding optical filaments dominated by forbidden-line emission, possibly indicative of shock excitation. In addition, we detect diffuse thermal X-ray emission inside of the shell possibly associated with it. 

\subsection{Possible SNR candidates (uncertain nature)}

\textbf{G319.8$-$2.0.} 
We report the detection of X-ray emission from the possible SNR candidate \blob.
While the source was previously noted in ROSAT data, its spectral properties could not be constrained. The improved sensitivity and energy coverage of eROSITA now allow a first spectral characterization, yielding parameters consistent with thermal emission from shock-heated plasma. 
\blob partly overlaps with the radio SNR candidate G320.0$-$1.7 proposed by \citet{2025ApJ...988...75B}, but the size of the X-ray region ($\sim10'$) is much smaller than the size of the radio structure ($\sim26'$), making the physical connection unlikely. Further observations are required to understand the nature of this object.\\

\noindent\textbf{G321.8$-$1.1.} 
G321.9$-$1.1 was previously identified in radio observations, where it appears as a roughly circular shell. In the eROSITA data, we detect faint diffuse X-ray emission within its boundary. Inspection of recent EMU data \citep{2021PASA...38...46N} reveals a smaller, inner circular structure inside the radio shell. The position and size of this inner feature matches that of the diffuse X-ray emission, which initially motivated us to consider whether it could represent a distinct SNR candidate, here referred to as G321.8--1.1.
However, the current data do not allow us to determine whether this inner structure is physically separate from G321.9--1.1 or instead forms part of the same remnant. No radio spectral-index measurement is currently available for either the larger shell or the inner feature, and the X-ray emission requires deeper data.

\noindent\textbf{G320.5$-$1.8.} 
The southern shell projected within the boundaries of G320.6$-$1.6 appears to be distinct from the larger remnant. It has a well-defined optical/radio shell, softer thermal X-ray emission than G320.6$-$1.6, and contains the Wolf--Rayet star WR\,68 within its projected boundary. These properties suggest that G320.5$-$1.8 may be a separate object rather than a substructure of G320.6$-$1.6. Its association with a Wolf--Rayet bubble is a plausible interpretation, although a separate SNR origin cannot be excluded with the current data.\\

Finally, our study highlights the importance of properly accounting for Galactic Ridge X-ray Emission (GRXE) when analysing extended sources close to the Galactic plane. In the soft X-ray band, where diffuse thermal emission from the GRXE can contribute significantly, an incomplete treatment of this component may influence the derived spectral parameters, in particular the absorption column density and plasma temperature.

\begin{acknowledgements}
 AK acknowledges support from the International Max-Planck Research School on Astrophysics at the Ludwig-Maximilians University (IMPRS). AK thanks Ruggero Valli, Xueying Zheng, Silvia Mantovanini, Ekaterina Makarenko, and Martin G. F. Mayer for the helpful discussions and suggestions.
 W.B. acknowledges support from the Deutsche Forschungsgemeinschaft through the project BE 1649/11-1\& BE 1649/11-2 within the Research Unit FOR 2990 (eRO-STEP). WB thanks Luciano Nicastro from INAF Bologna for helpful discussions.
 This work is based on data from eROSITA, the soft X-ray instrument aboard \textit{SRG}, a joint Russian-German science mission supported by the Russian Space Agency (Roskosmos), in the interests of the Russian Academy of Sciences represented by its Space Research Institute (IKI), and the Deutsches Zentrum für Luft- und Raumfahrt (DLR). The \textit{SRG} spacecraft was built by Lavochkin Association (NPOL) and its subcontractors, and is operated by NPOL with support from the Max Planck Institute for Extraterrestrial Physics (MPE). The development and construction of the eROSITA X-ray instrument was led by MPE, with contributions from the Dr. Karl Remeis Observatory Bamberg \& ECAP (FAU Erlangen-Nuernberg), the University of Hamburg Observatory, the Leibniz Institute for Astrophysics Potsdam (AIP) and the Institute for Astronomy and Astrophysics of the University of Tübingen, with the support of DLR and the Max Planck Society. The Argelander Institute for Astronomy of the University of Bonn and the Ludwig Maximilians Universität Munich also participated in the science preparation for eROSITA. 
 The eROSITA data shown here were processed using the eSASS software system developed by the German eROSITA consortium. 
 This research has made use of the Digitized Sky Survey, which were produced at the Space Telescope Science Institute under U.S. Government grant NAG W-2166.
 The images of these surveys are based on photographic data obtained using the Oschin Schmidt Telescope on Palomar Mountain and the UK Schmidt Telescope. The plates were processed
 into the present compressed digital form with the permission of these institutions.
\end{acknowledgements}

\bibliographystyle{aa}
\bibliography{bib}

\begin{appendix}
\label{sec:appendix}

\section{Point source removal}

To estimate the systematic uncertainty introduced by our point source removal procedure, we evaluated the contribution of faint sources that remain in the data. In our analysis we adopted a conservative threshold of $\mathrm{DET\_LIKE\_0} > 30$ for point-source removal. Sources below this threshold were not masked and may therefore contribute residual counts within the extended source extraction regions.

To quantify this effect, we over-plotted sources with $6 < \mathrm{DET\_LIKE} < 30$ on the adaptively smoothed image and visually inspected them to identify likely spurious detections. Sources that appeared to be artifacts of diffuse emission or background fluctuations were excluded. The lower threshold of $\mathrm{DET\_LIKE}=6$ was adopted because it corresponds approximately to a $3\sigma$ detection significance and is the threshold used in the construction of the main eROSITA source catalogue.

We then summed the counts from the remaining sources, obtaining a total of 140 counts. Compared to the total number of counts measured in this region ($13507$), this corresponds to a fractional contribution of approximately $1\%$. We therefore conclude that the residual contamination from real point sources with $\mathrm{DET\_LIKE} < 30$ has a negligible impact on our spectral analysis.

\end{appendix}

\end{document}